\documentclass{mn2e}
\usepackage{graphicx}

\newcommand{\etal}{et~al.}
\newcommand{\ionhy}{H{\sc ii}}
\newcommand{\UCHII}{UCH{\sc ii}}

\newcommand{\kms}{$\mbox{km~s}^{-1}$}
\newcommand{\cc}{$\mbox{cm}^{-3}$}
\newcommand{\pccm}{$\mbox{pc~cm}^{-6}$}
\newcommand{\otherpaper}{ESK05}

\begin{document}

\title[\ionhy\/ Regions in Hierarchical Clouds] 
{Evolution of \ionhy\/ regions in hierarchically structured molecular clouds}

\author[Shabala, Ellingsen, Kurtz \& Forbes]{S.S. Shabala$^{1,2}$, S.P. Ellingsen$^{1\star}$, 
S.E. Kurtz$^3$, L.K. Forbes$^1$\\
$^1$ School of Mathematics and Physics, University of Tasmania, 
Private Bag 21, Hobart, Tasmania 7001, Australia\\
$^2$ Astrophysics Group, Cavendish Laboratory, Madingley Road, Cambridge CB3 0HE, United Kingdom\\
$^3$ Centro de Radioastronom\'\i a y Astrof\'\i sica, UNAM,
Apdo Postal 3-72, CP 58089, Morelia, Michoac\'an, M\'exico\\
Email: Simon.Ellingsen@utas.edu.au}

\maketitle

\begin{abstract}

  We present observations of the H\,91$\alpha$ recombination line
  emission towards a sample of nine \ionhy\/ regions associated with
  6.7-GHz methanol masers, and report arcsecond-scale emission around
  compact cores. We derive physical parameters for our sources, and
  find that although simple hydrostatic models of region evolution
  reproduce the observed region sizes, they significantly
  underestimate emission measures. We argue that these findings are
  consistent with young source ages in our sample, and can be
  explained by existence of density gradients in the ionised gas.

\end{abstract}


\section{Introduction} \label{sec:introduction}

Ultracompact (UC) \ionhy\/ regions are pockets of ionised hydrogen
that form around massive stars in the earliest stages of their
evolution. Together with massive bipolar outflows, strong far infrared
emission by dust, and the presence of molecular masers, they are
indicative of massive star formation in its earliest stages. The
simple model of \ionhy\/ region evolution \cite{Spitzer78} does not
explain many of their observed properties, such as the frequent
occurrence of non-spherical morphologies and the lifetime paradox.
This lifetime problem, first noted by Wood \& Churchwell
\shortcite{WoodChurchwell89a}, is that the number of observed UC
\ionhy\/ regions exceeds that predicted from their dynamical expansion
timescales by two orders of magnitude, given the accepted massive star
formation rate in the Galaxy.  A number of modifications and
enhancements to the basic model have been suggested, including the
work of Dyson \etal\ \shortcite{DysonEA95}, Hollenbach \etal\
\shortcite{HollenbachEA94}, Tenorio-Tagle \shortcite{Tenorio-Tagle79}
and van~Buren \etal\ \shortcite{vanBurenEA90}, Franco \etal\
\shortcite{FrancoEA90}, Arthur \& Lizano \shortcite{Arthur97},
Keto~\shortcite{Keto03}.  The thermal \cite{DePreeEA95a} and turbulent
\cite{XieEA96} pressure confinement models are appealing due to their
dependence on the ambient conditions observed to commonly exist in
molecular clouds.

De~Pree \etal\/ \shortcite{DePreeEA95a} suggested thermal pressure
confinement as an explanation of the lifetime paradox.  They noted
that when Wood \& Churchwell proposed the lifetime problem in 1989,
the molecular medium surrounding the UC \ionhy\/ regions was thought
to have temperatures $\sim 25$~K and densities $\sim 10^5$~cm$^{-3}$.
More recent observations indicate T$\sim 100$~K and n$\sim
10^7$~cm$^{-3}$.  The resulting 400$\times$ increase in thermal
pressure limits the expansion of the Str\"omgren sphere.  A weakness
of this model, noted by Xie \etal\/ \shortcite{XieEA96}, is the
exceedingly high emission measures that it predicts ($\sim
10^{10}$~\pccm), which are more than two orders of magnitude greater
than the values typically observed.

Hierarchical density and temperature structures are known to exist
within star-forming regions, with hot cores embedded in larger, less
dense molecular clumps which themselves are within still larger and
less dense molecular clouds. The densities decrease by approximately
an order of magnitude in going from core to clump and again from clump
to cloud \cite{CesaroniEA94}.  In a seminal work, Franco et al.
\shortcite{FrancoEA90} showed that these density inhomogeneities are
important for \ionhy\/ region evolution. That the hierarchical
structure of molecular clouds plays an important role in \ionhy\/
evolution is supported by the marked similarity in the ionised and
neutral gas density structures observed within molecular clouds
\cite{KimKoo96,KimKoo02,KooKim03}.

The large thermal molecular line and recombination line widths
observed towards many UC \ionhy\/ regions suggest that significant
turbulent motions are present, probably of magnetic origin
\cite{GarciaSeguraFranco96}. This led Xie \etal\/ \shortcite{XieEA96}
to suggest that turbulent pressure is the dominant mechanism to
restrict the expansion of an \ionhy\/ region. In contrast to thermal
pressure confinement, the assumed densities are lower, resulting also
in lower emission measures. During much of the expansion phase,
however, the turbulent pressure is expected to play a lesser role than
thermal pressure in confining the \ionhy\/ region. This is discussed
briefly in section~\ref{sec:turb}.

Icke \shortcite{Icke79} investigated the formation of \ionhy\/ regions
in non-homogeneous media and was able to explain some non-spherical
morphologies.  Observational evidence obtained in the last decade,
however, suggests that many \ionhy\/ regions have compact cores within
diffuse, arcminute-scale extended emission \cite{KurtzEA99,KimKoo01}.
Other sources exhibit this to a smaller degree --- the so-called 
core-halo morphology; see Wood \& Churchwell~\shortcite{WoodChurchwell89a} and 
Kurtz \etal\/~\shortcite{KurtzEA94}.  A study of the compact and extended radio
continuum emission and radio recombination lines (RRLs) from eight
\ionhy\/ regions known to be associated with 6.7-GHz methanol masers
has been undertaken. The RRL analysis is presented here, while details
of the continuum observations can be found in Ellingsen, Shabala \&
Kurtz \shortcite{EllingsenEA05} (hereafter~\otherpaper). In
section~\ref{sec:observations} we briefly outline our
observations. The results are presented in section~\ref{sec:results},
and these are compared with a simple model in
section~\ref{sec:model}. A discussion of our findings is presented in
section~\ref{sec:discussion}.

\section{Observations}
\label{sec:observations}

Eight UC \ionhy\/ regions with associated methanol maser emission were
observed with the Australia Telescope Compact Array (ATCA) on 1999
July 10 and 11. Both continuum and recombination line emission were
observed in the ATCA 750D array, with an angular resolution of $7''$
and largest detectable angular scale of $\sim 50''$.  Details of the
continuum observations are found in \otherpaper\/.

The H91$\alpha$ recombination line ($\nu_0 = 8.58482$~GHz) was
observed with an 8-MHz bandwidth and 512 spectral channels, giving a
frequency resolution of 15.625~kHz (0.529~km~s$^{-1}$), and total
velocity coverage of 270~km~s$^{-1}$. The data were further smoothed
by frequency averaging over four or eight channels. All sources were
observed together with associated secondary calibrators immediately
before and after each on-source observation. The primary calibrator
PKS\,B1934-638 was observed each day to calibrate the flux density
scale.

The observations were made with the array in the 750D configuration,
with a minimum baseline of 31~m and maximum of 719~m. Because the
primary aim of these observations was to look for extended emission
associated with the sources, baselines to the 6~km antenna were not
used in order to maximize sensitivity to large-scale structure.  A
summary of the fields observed is given in Table~1.

\begin{table*}
        \begin{centering}
        \begin{tabular}{|l|c|c|c|c|c|c}
        \hline
        Source & {\small{Methanol peak}} & Right Ascension & Declination  & Associated   &  {\small{Central V$_{LSR}$}} & Avg noise per spectral channel \\
               & {\small{(km s$^{-1}$)}} & (J2000)        & (J2000)      & IRAS Source & {\small{(km s$^{-1}$)}} & (mJy beam$^{-1}$)\\
        \hline
        G\,308.92+0.12 & -54.5 $^a$ & 13:43:02 & -62:08:51 & 13395-6153 &  21.8 & 2.3\\
        G\,309.92+0.48 & -59.6 $^b$ & 13:50:42 & -61:35:10 & 13471-6120 &  21.6 & 2.0\\
        G\,318.95-0.20 & -34.7 $^b$ & 15:00:55 & -58:58:42 & 14567-5846 &  18.3 & 3.1\\
        G\,328.81+0.63 & -44.0 $^c$ & 15:55:48 & -52:43:07 & 15520-5234 &  14.7 & 3.9\\
        G\,336.40-0.25 & -85.3 $^a$ & 16:34:11 & -48:06:26 & none       &  11.2 & 3.3 \\
        G\,339.88-1.26 & -38.7 $^d$ & 16:52:05 & -46:08:34 & 16484-4603 &  9.3 & 2.8 \\
        G\,345.01+1.79 & -18.0 $^d$ & 16:56:48 & -40:14:26 & 16533-4009 &  7.8 & 4.1 \\
        NGC\,6334F & -10.4 $^a$     & 17:20:53 & -35:47:01 & 17175-3544 &  4.4 & 5.5 \\
        NGC\,6334E & - $^e$     & 17:20:53 & -35:47:01 & 17175-3544 &  4.4 & 5.5\\
        \hline
        \end{tabular}
        \end{centering}
        \caption{Observed Sources. {\itshape{a)}} Phillips et
          al. \protect\shortcite{PhillipsEA98}; {\itshape{b)}} Walsh et
          al.  \protect\shortcite{WalshEA98}; {\itshape{c)}} Ellingsen
          et al. \protect\shortcite{EllingsenEA03}; {\itshape{d)}}
          Ellingsen et al. \protect\shortcite{EllingsenEA96}. {\itshape{e)}} 
          Observed within the same primary beam as NGC 6334F.  
          Tabulated right ascension and declination values refer to the 
          pointing centre (usually the methanol maser site), which is
          usually close to, but not necessarily coincident with the HII region
          centre.}
        \label{tab:ObservedSources}
\end{table*}

\section{Results}
\label{sec:results}

\begin{figure*}
        \centering
        \includegraphics[height=0.3\textwidth]{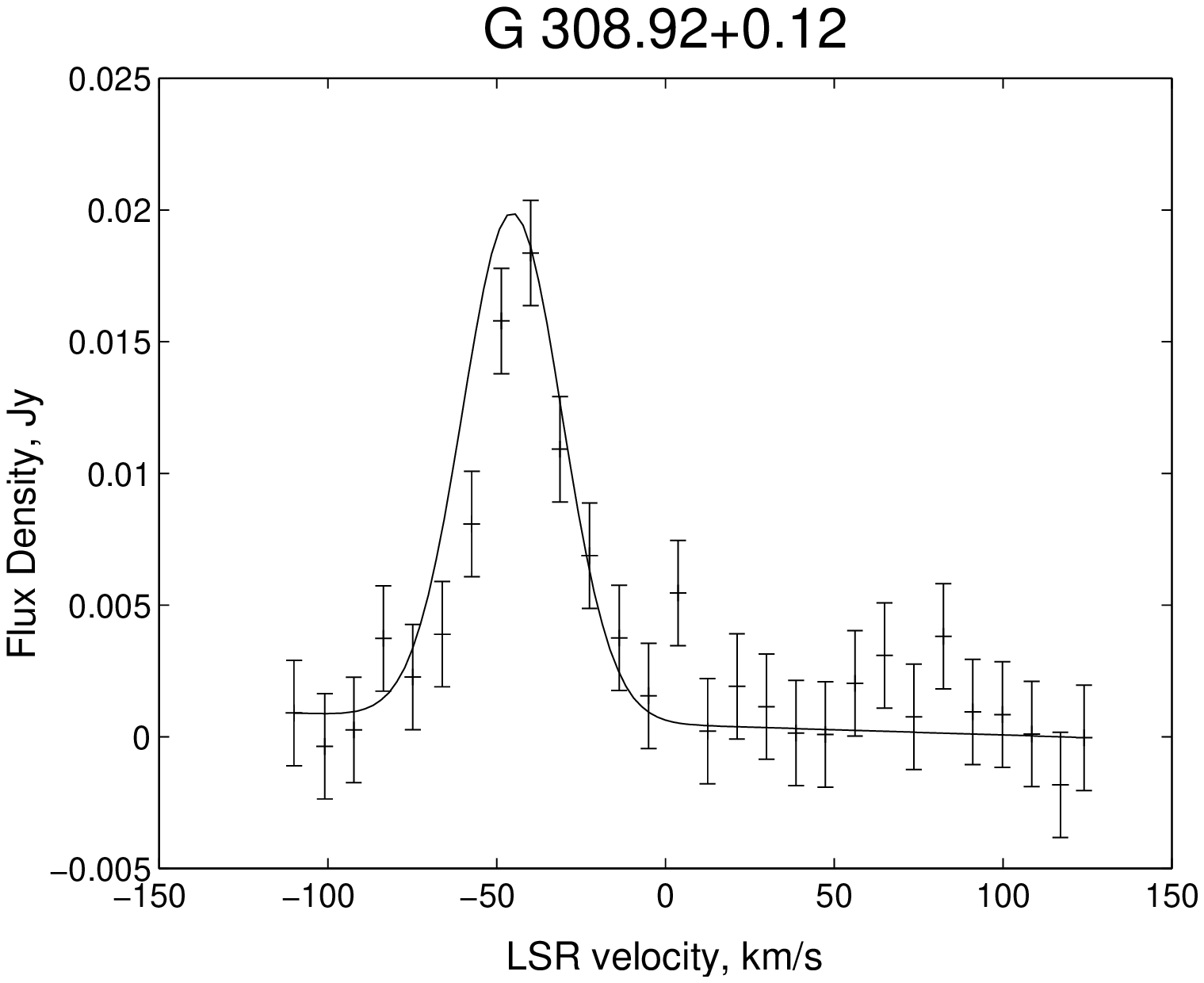}
        \includegraphics[height=0.3\textwidth]{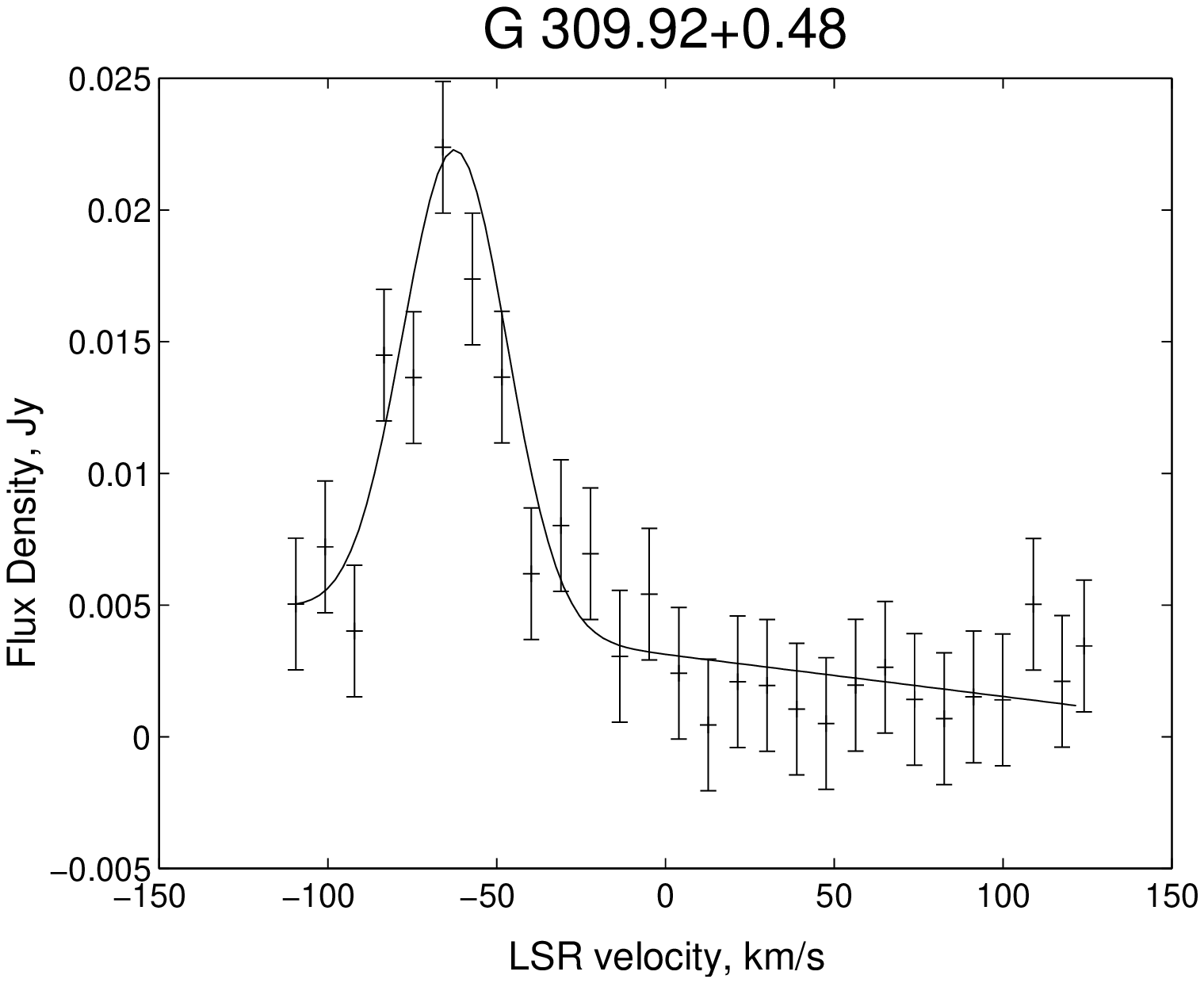}
        \includegraphics[height=0.3\textwidth]{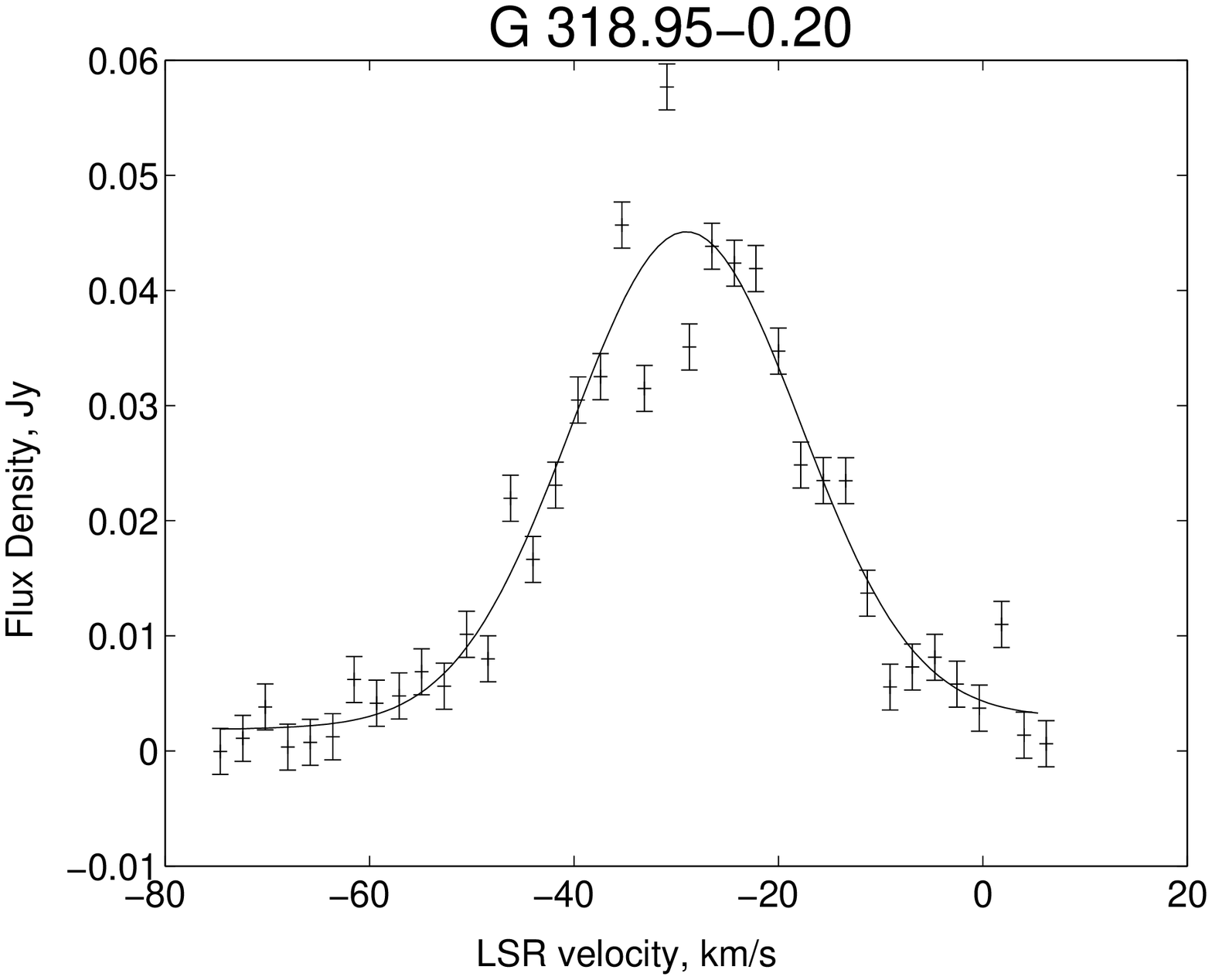}
        \includegraphics[height=0.3\textwidth]{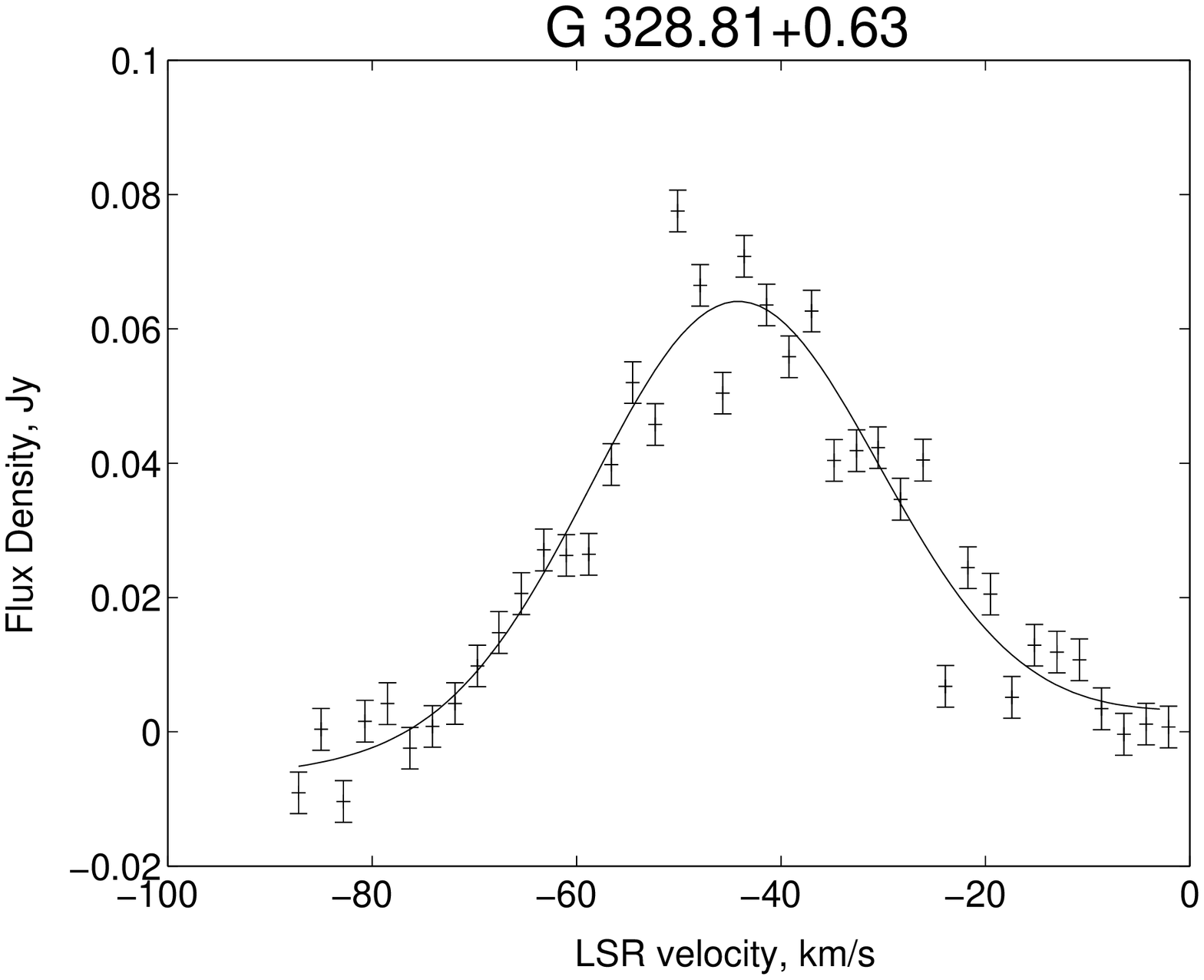}
        \includegraphics[height=0.3\textwidth]{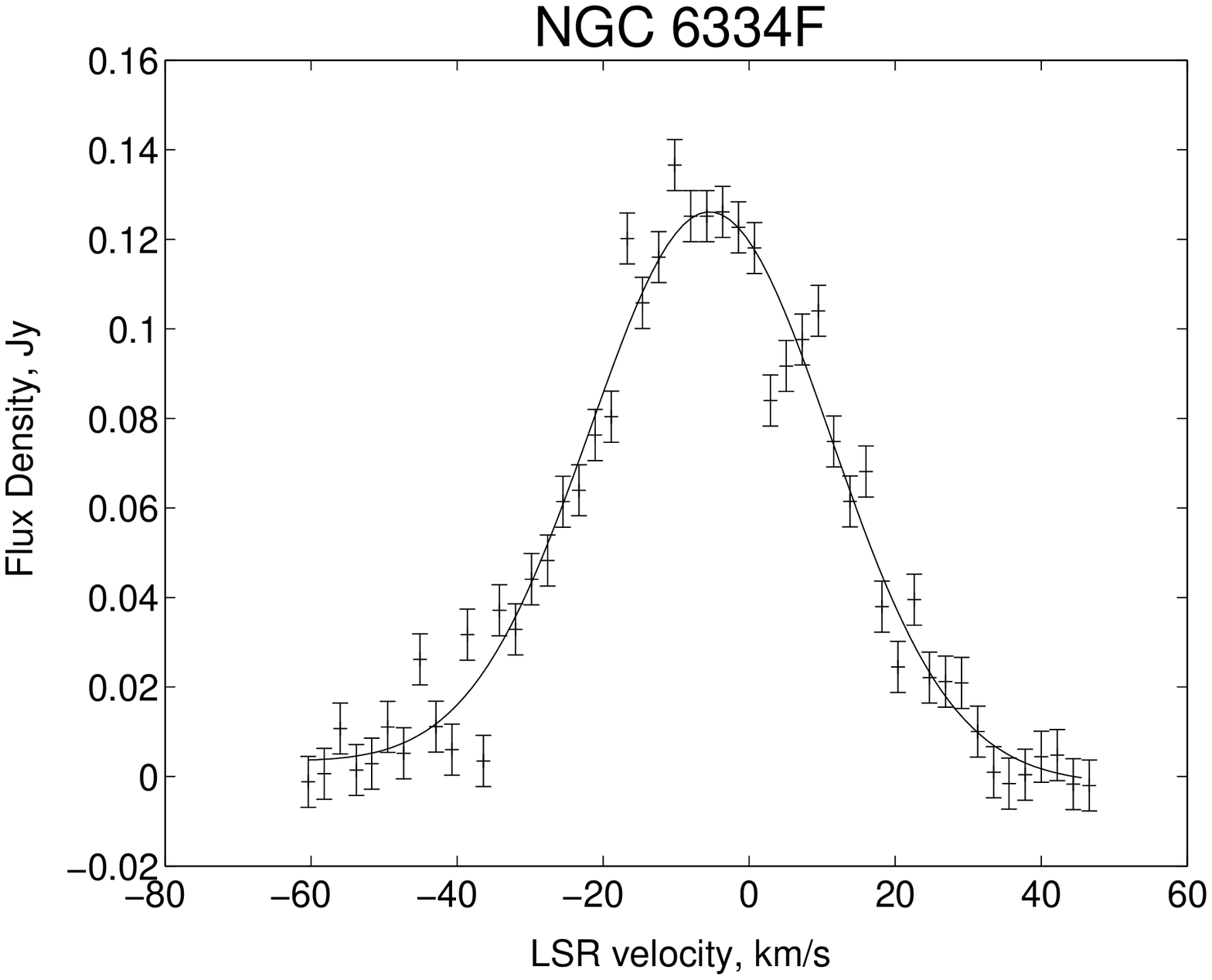}
        \includegraphics[height=0.3\textwidth]{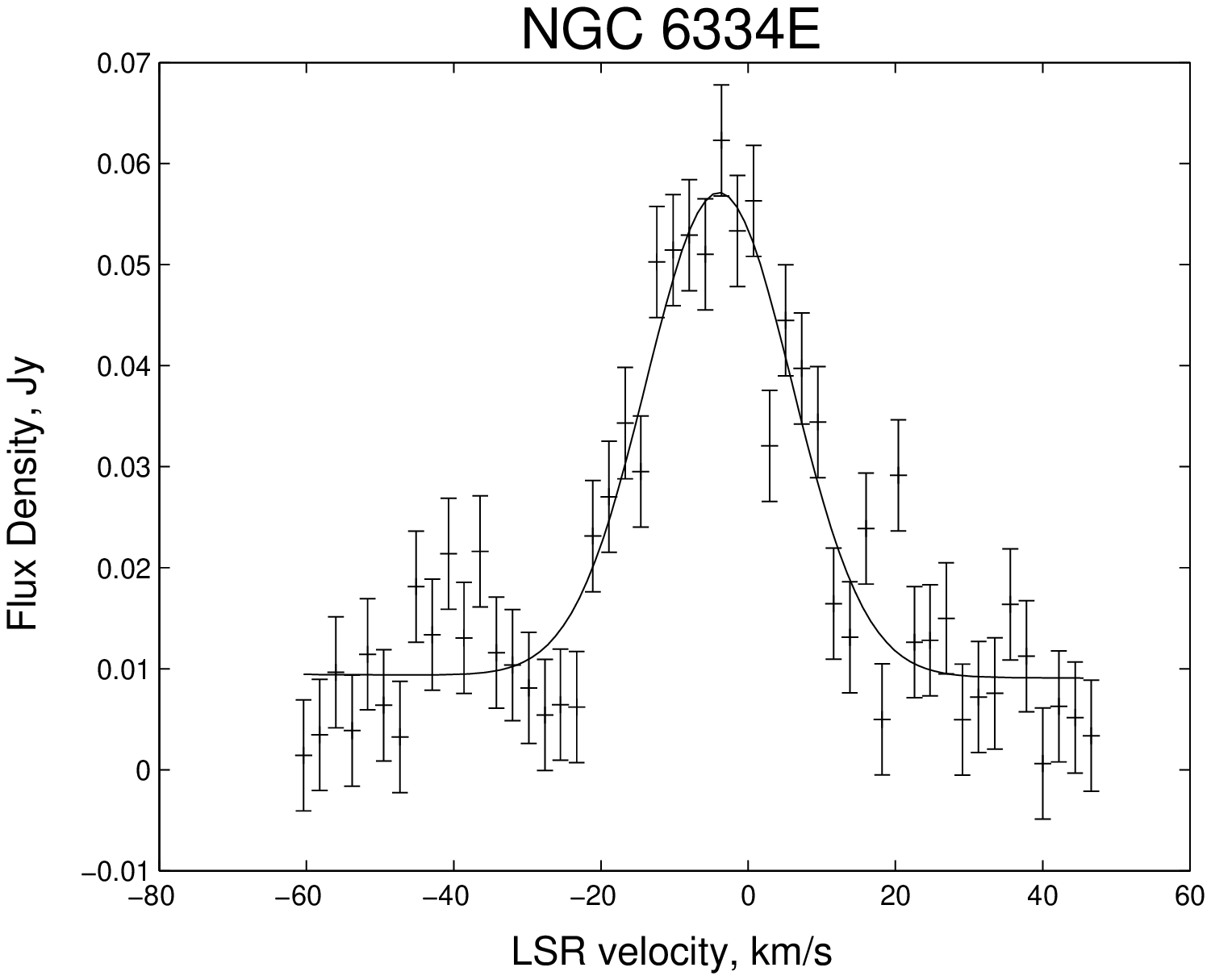}
        \caption{Gaussian least-squares fits to recombination line data. Error bars indicate the average noise level in each channel. Detection for source G309.92+0.48 is marginal, and a best-fit Gaussian is shown only for illustrative purposes.}
        \label{fig:RRLs}
\end{figure*}

H91$\alpha$ emission was detected in six sources. The
observed line profiles and Gaussian fits obtained after continuum subtraction
are shown in figure~\ref{fig:RRLs}. As we are only interested in emission
around the RRL peak, non-zero baselines were used in some sources for a
better fit in those regions. The flux densities per channel were obtained
by integrating over the area of the UC \ionhy\/ region (typically around $5''
\times 5''$; see Table~\ref{tab:GaussianParameters}). No H91$\alpha$ emission was
detected from G336.40$-$0.25, G339.88$-$1.26 or G345.01+1.79.  
These three regions have low continuum
brightness so the expected LTE line brightness is at or below the
image noise level.

The fit parameters are given in Table~\ref{tab:GaussianParameters}.
These were achieved using a standard non-linear least squares fitting
routine, and resulted in unrealistically low error estimates for the
fit parameters. In order to determine a realistic uncertainty, we made
Monte-Carlo simulations, using the same routine to fit Gaussian
profiles with known parameters, plus white noise of various
amplitudes.  Comparison of the methanol maser peak velocities from
Table~\ref{tab:ObservedSources} with the H91$\alpha$ velocities from
Table~\ref{tab:GaussianParameters} shows approximate agreement
(indicating an association between the masers and the star formation
region) but sufficient difference in some cases to suggest that the
masers may not be directly linked to the ionised gas.  The detection
is only marginal for G309.92+0.48 and thus no formal parameters were
derived for this source.

\begin{table*}
  \begin{centering}
  \begin{tabular}{|l|c|c|c|c|c|c|c|c|c}
  \hline
  Source & {\small{V$_{peak}$}} & {\small{$S_L$}} & {\small{FWHM}} & {\small{$S_C$}} & {\small{$\Omega_L$}} & {\small{$\Omega_C$}} & {\small{$T_C$}} & {\small{$T_e$}}  & {\small{source size}} \\
  & {\small{(km s$^{-1}$)}} & {\tiny{(mJy)}} & {\small{(km s$^{-1}$)}} & {\tiny{(Jy)}} & {\small{(arcsec)}} & {\small{(arcsec)}} & {\small{(K)}} & {\small{(K)}} & {\small{(arcsec)}}\\
  \hline
  G\,308.92+0.12  & -44.4 $\pm$ 0.8 &  19.9 $\pm$ 1.3 & 34.7 $\pm$ 4.6 & 0.263 $\pm$ 0.0002 & $7.4 \times 6.1$ & $7.3 \times 6.8$ & 3900 $\pm$ 20 & 8200 $\pm$ 1400 & $8.6 \times 4$.0\\
  G\,309.92+0.48  &  - & - & - & 0.670 $\pm$ 0.0003 & $7.6 \times 6.3$ & $7.4 \times 7.0$ & - & - & $8.4 \times 4.0$\\
  G\,318.95-0.20  & -29.0 $\pm$ 0.4   &  45.1 $\pm$ 0.7 & 26.4 $\pm$ 1.4 & 0.724 $\pm$ 0.001  & $8.1 \times 6.4$ & $8.1 \times 6.8$ & 9630 $\pm$ 40 & 12600 $\pm$ 800 & $7.7 \times 4.0$\\
  G\,328.81+0.63  & -44.5 $\pm$ 0.3   &  64.1 $\pm$ 1.2 & 33.2 $\pm$ 0.9 & 1.47  $\pm$ 0.0005  & $8.2 \times 6.1$ & $8.5 \times 7.0$ & 18370 $\pm$ 80 & 12900 $\pm$ 500 & $6.6 \times 4.0$\\
  NGC\,6334F      & -5.3  $\pm$ 0.2 & 126.1 $\pm$ 1.6 & 38.8 $\pm$ 0.8 & 2.13  $\pm$ 0.06   & $11.7 \times 7.4$ & $11.3 \times 7.6$ & 18300 $\pm$ 600 & 10100 $\pm$ 500 & $5.0 \times 4.0$\\
  NGC\,6334E      & -4.0  $\pm$ 0.2 &  57.1 $\pm$ 2.1 & 23.4 $\pm$ 1.3 & 0.417 $\pm$ 0.002  & $11.7 \times 7.4$ & $11.3 \times 7.6$ & 3580 $\pm$ 30 & 7500 $\pm$ 600 & $5.0 \times 4.0$\\
  \hline
  \end{tabular}
  \end{centering}
  \caption{H91$\alpha$ non-linear least squares Gaussian fits and
  derived parameters. Source sizes are FWZI. $S_C$ values are taken
  from \otherpaper. Errors for the fits are estimated from Monte Carlo
  simulations (see text). V$_{peak}$, $S_L$, and FWHM are obtained from the
  Gaussian fits. The electron temperature, $T_e$ and continuum brightness
  temperature, $T_C$ are derived as outlined in 
  Section~\ref{sec:ElectronTemps}. For G\,309.92+0.48 we have only a marginal
  detection, and hence no formal fit; as a consequence no $S_L$ is 
  listed and no $T_e$ estimate can be made.}
        \label{tab:GaussianParameters}
\end{table*}

\subsection{Arcsecond-scale emission}
\subsubsection{Moment maps}
\label{sub:MomentOutline}

Image cubes for the six sources that exhibited H91$\alpha$
emission were analysed for the presence of arcsecond-scale extended emission near the
UC \ionhy\/ regions. Two complementary methods were employed
for this analysis.

\begin{figure*}
        \centering
        \includegraphics[height=0.4\textwidth]{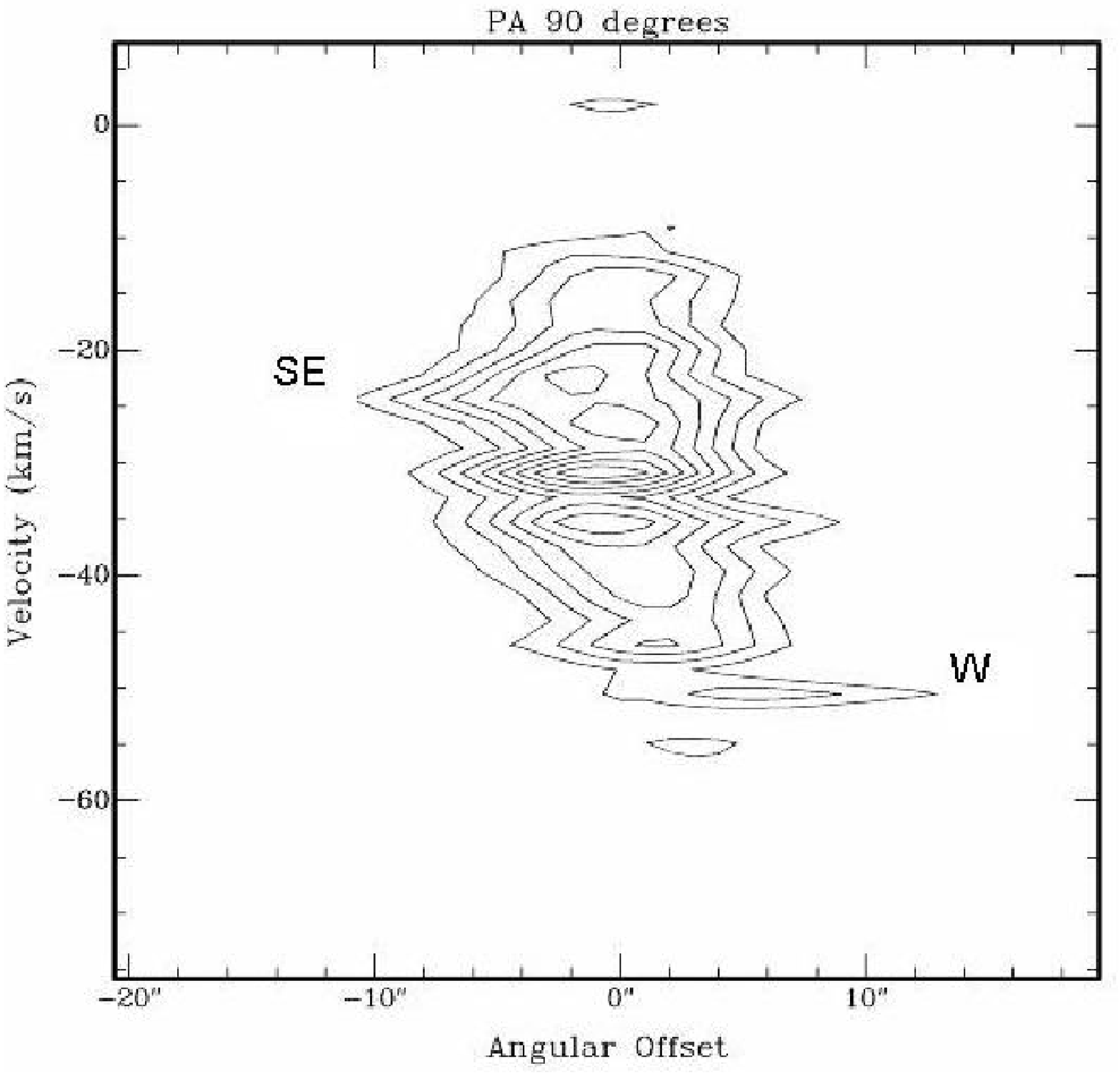}
        \includegraphics[height=0.4\textwidth]{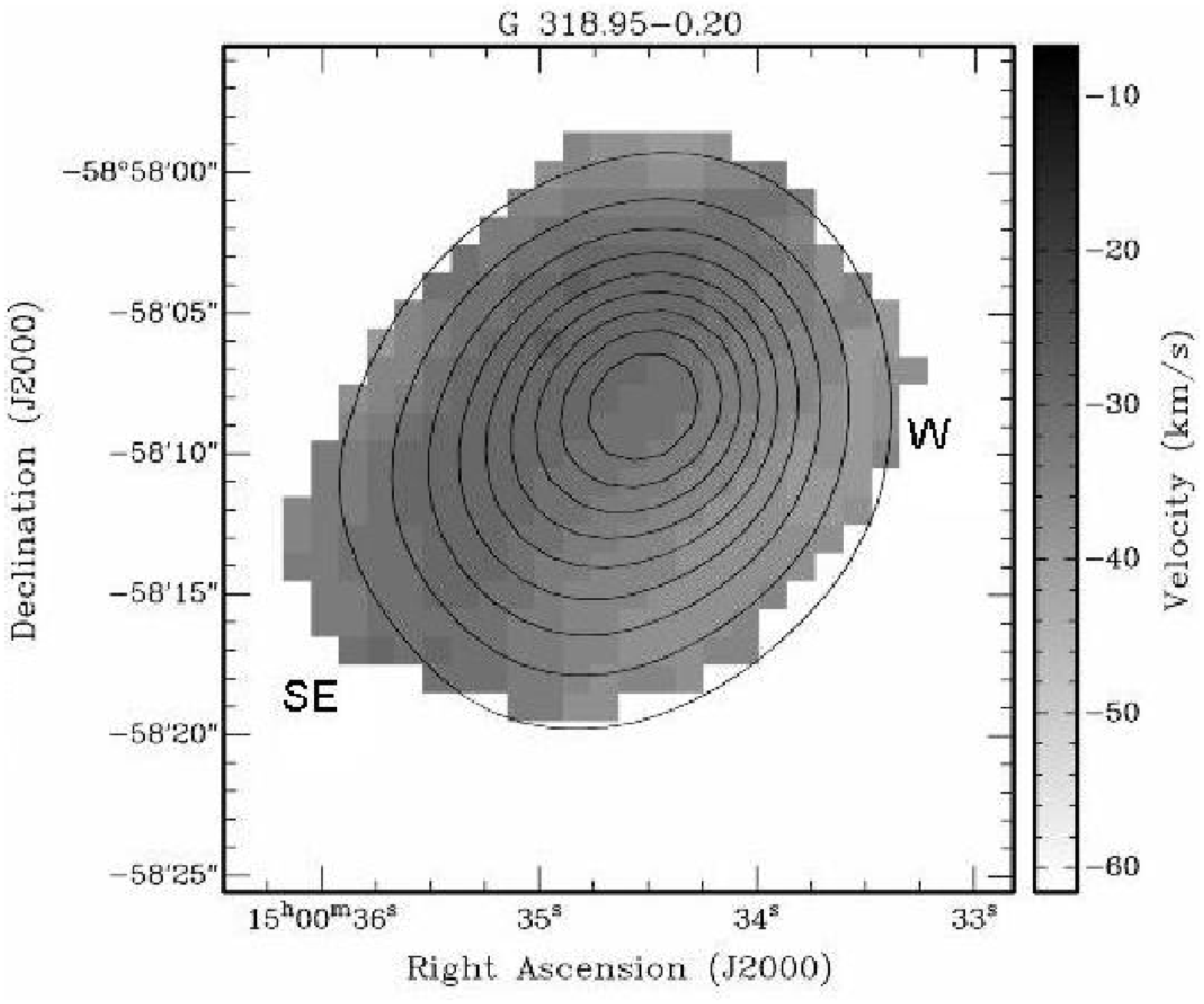}
        \caption{H91$\alpha$ moment maps for G318.95$-$0.20. 
Position-velocity plots are shown in part~{\itshape{a}} (left plot). 
Contours are at 20 to 90\% of maximum emission of 
45.1~mJy~beam$^{-1}$ spaced by 10\%. The lowest contour of 9.0~mJy~beam$^{-1}$ corresponds
to $2.9 \sigma$. Velocity distribution across the 
region is shown in grey scale in part~{\itshape{b}} (right plot). 
Also plotted in part~{\itshape{b}} are continuum contours at 10 to 
90\% of maximum emission of 0.724~Jy~beam$^{-1}$ spaced by 10\%. 
From the 90$^\circ$ cross-cut (corresponding to a cut in RA), the slight western
extension seems to move at a different characteristic velocity and is
therefore probably not associated with other components, or is perhaps
related to an outflow. A south-eastern extension is consistent with
observed elongation in that direction on the continuum plot, thus the
compact emission and an extension to the south-east are likely to be
associated. Velocity slices suggest the amount of extended emission around this source is
larger than for the others. The continuum and RRL emission are 2$'$
distant from the methanol maser, suggesting that this region may be older
than the other sources in our sample.}
        \label{fig:Moment318}
\end{figure*}

\begin{figure*}
        \centering
        \includegraphics[height=0.45\textwidth]{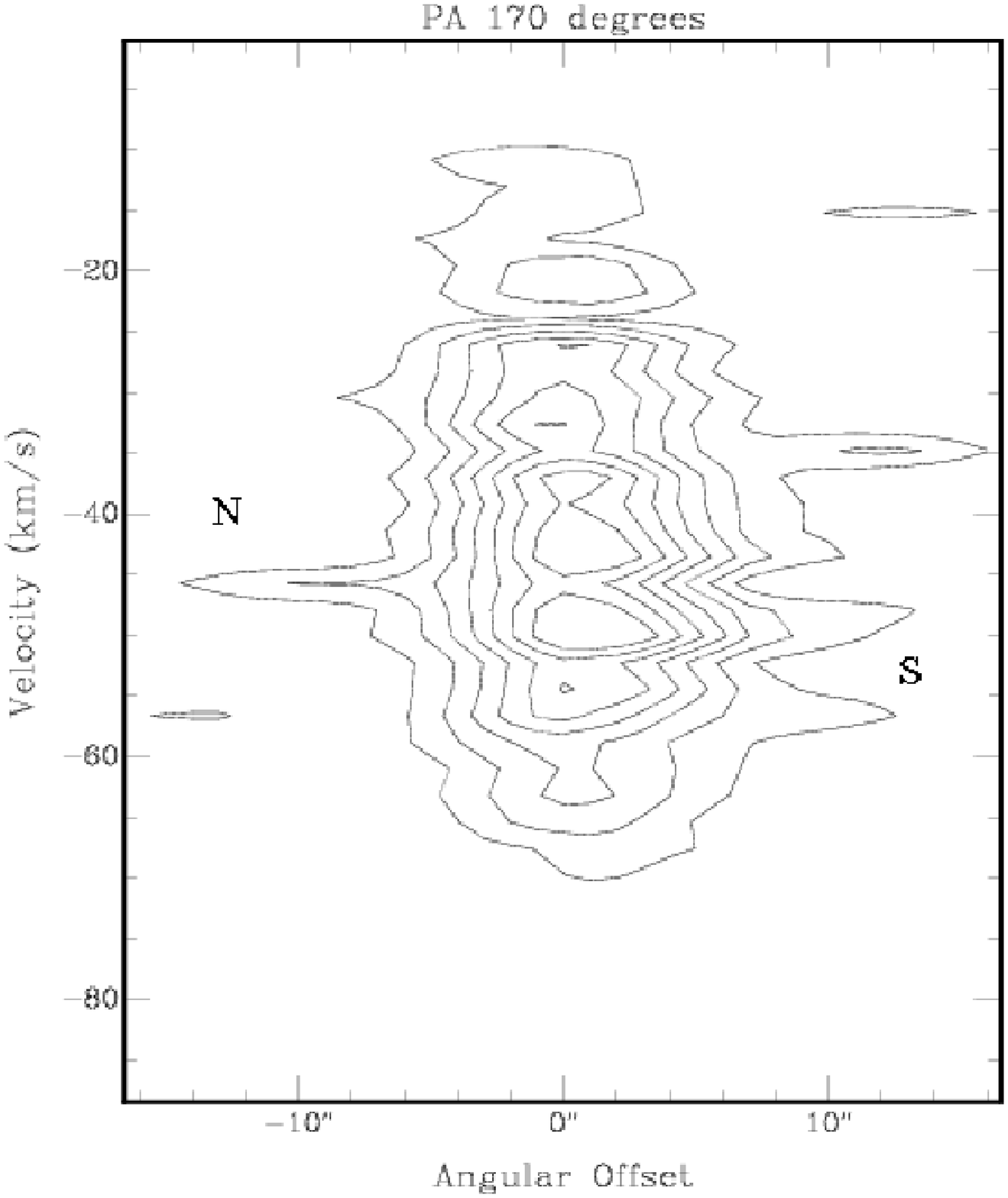}
        \includegraphics[height=0.45\textwidth]{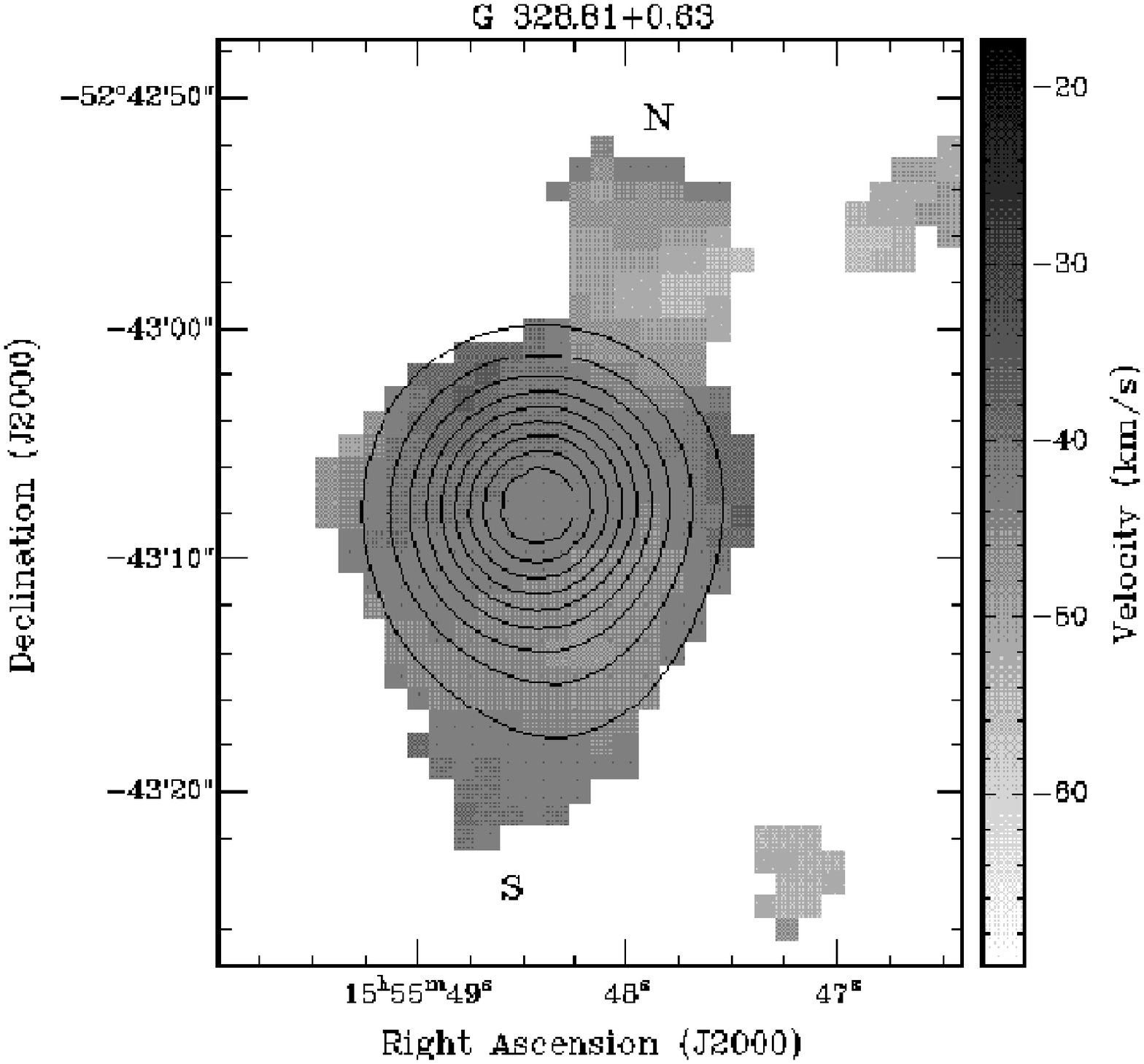}
        \caption{H91$\alpha$ moment maps for G328.81+0.63. 
The plots are as in Figure~\ref{fig:Moment318}. Position-velocity 
contours are at 25 to 85\% of maximum emission of 64.1 mJy beam$^{-1}$ 
spaced by 10\%, with the lowest contour of 16.0~mJy~beam$^{-1}$ corresponding
to $4.1 \sigma$. Continuum contours are at 10 to 90\% of peak 
emission of 1.46~Jy~beam$^{-1}$ spaced by 10\%. Two distinct components
are seen to the north and south of the compact core. Both
are found to move at velocities close to those of the core, and are
connected in the velocity slices. Hence the extended emission
in this source is likely to be associated with the compact component.
The north component coincides with the peak continuum
emission observed in the continuum map, while the
south extension corresponds to the cometary tail (see Figure~4 of \otherpaper). 
An additional north-west component separate from the rest of the emission also
appears at approximately -35~km~s$^{-1}$.}
        \label{fig:Moment328}
\end{figure*}

\begin{figure*}
   \centering
   \includegraphics[height=0.35\textwidth]{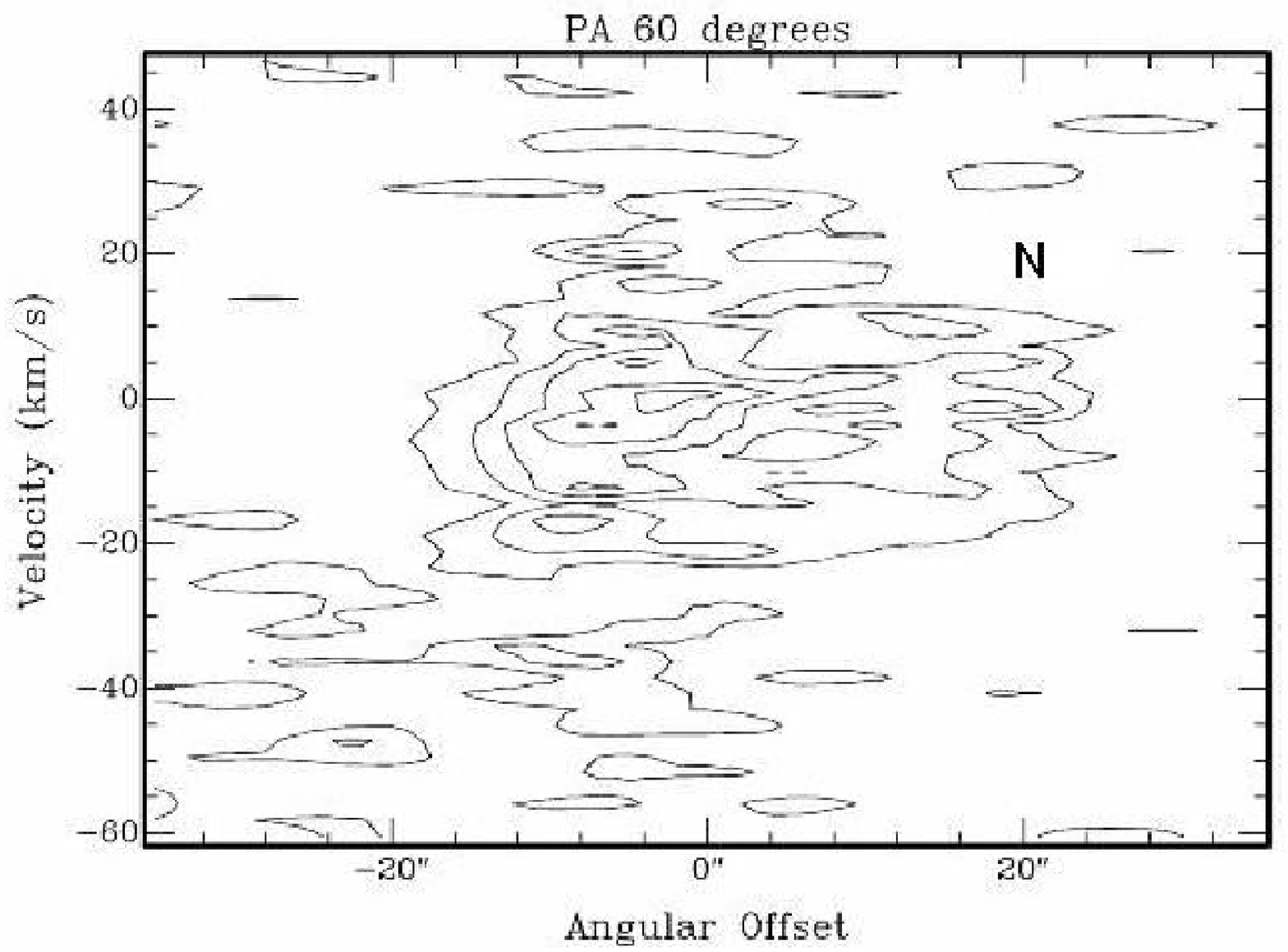}
   \includegraphics[height=0.35\textwidth]{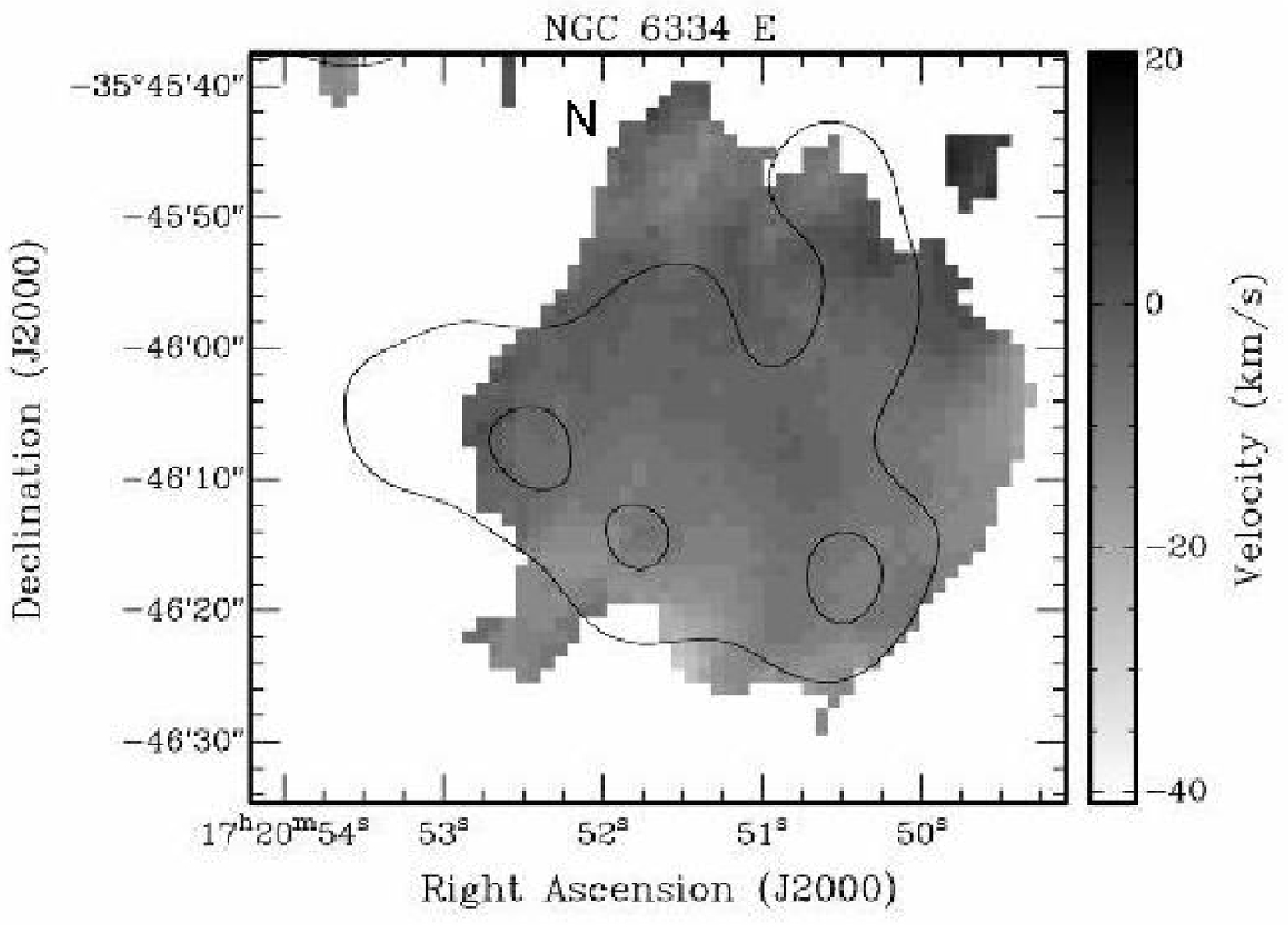}
   \caption{H91$\alpha$ moment maps for NGC~6334~E. The plots are as
            in Figure~\ref{fig:Moment318}. Position-velocity contours
            are at 35 to 95\% of maximum emission of 57.3 mJy
            beam$^{-1}$ spaced by 15\%. The lowest contour of
            20.1~mJy~beam$^{-1}$ corresponds to $3.6
            \sigma$. Continuum contours are at 82 and 90\% of peak
            emission. The extended and central components of the RRL
            emission, separated by about 20 arcseconds on the sky, are
            practically coincident in the 60$^\circ$ cut shown. All
            cross-cuts suggest these move at a very similar velocity,
            and it is thus likely that they are associated. The level
            of background continuum emission is rather high in this
            source, resulting in poor image and p-v plot quality
            compared with other sources. High sensitivity VLA
            observations show NGC 6334 E to have a shell-like
            morphology with a weak, central compact source
            \protect\cite{CarralEA02}. This structure is not well-imaged in
            the ATCA observations presented here.}
   \label{fig:MomentNGCe}
\end{figure*}

\begin{figure*}
   \centering
   \includegraphics[height=0.35\textwidth]{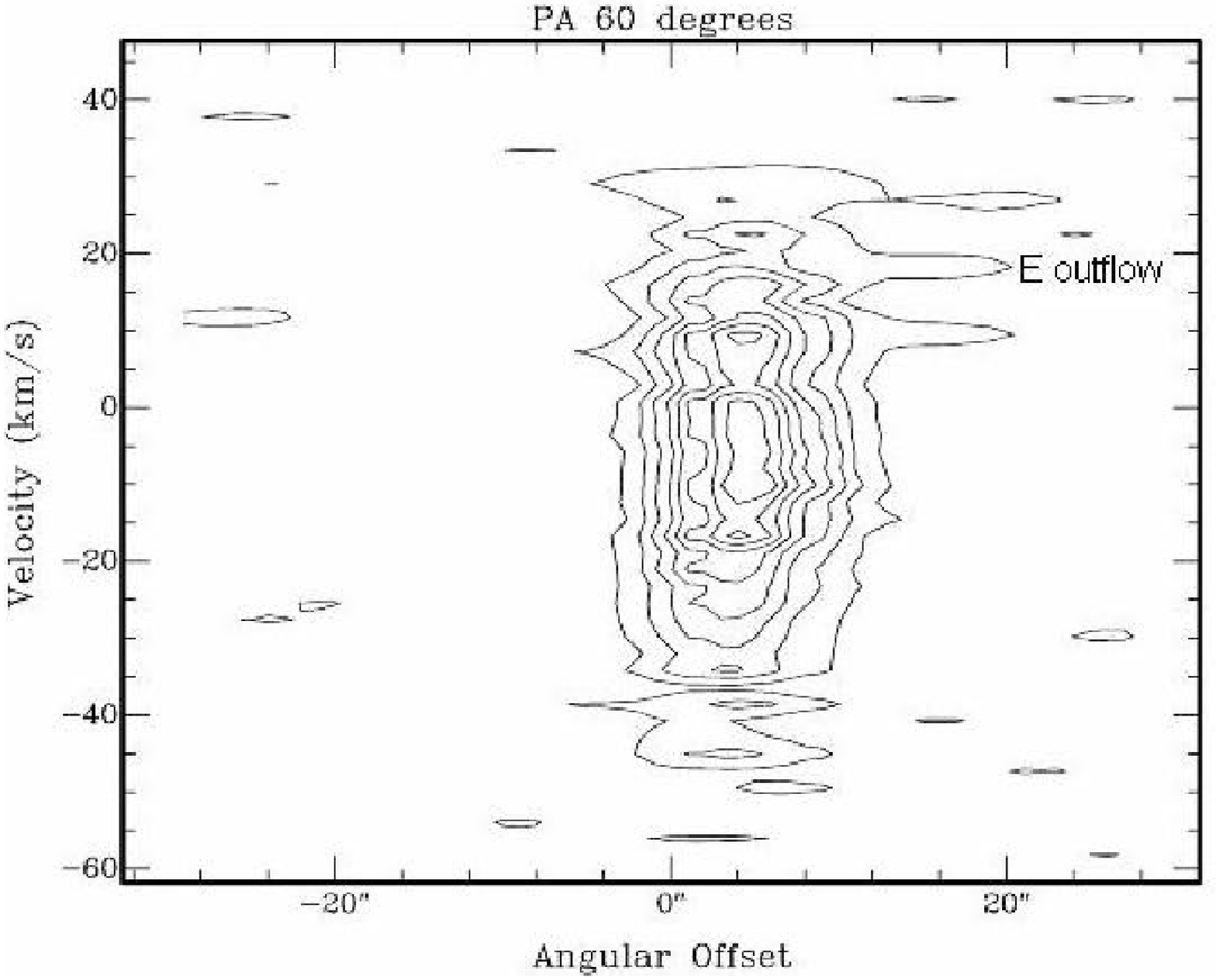}
   \includegraphics[height=0.35\textwidth]{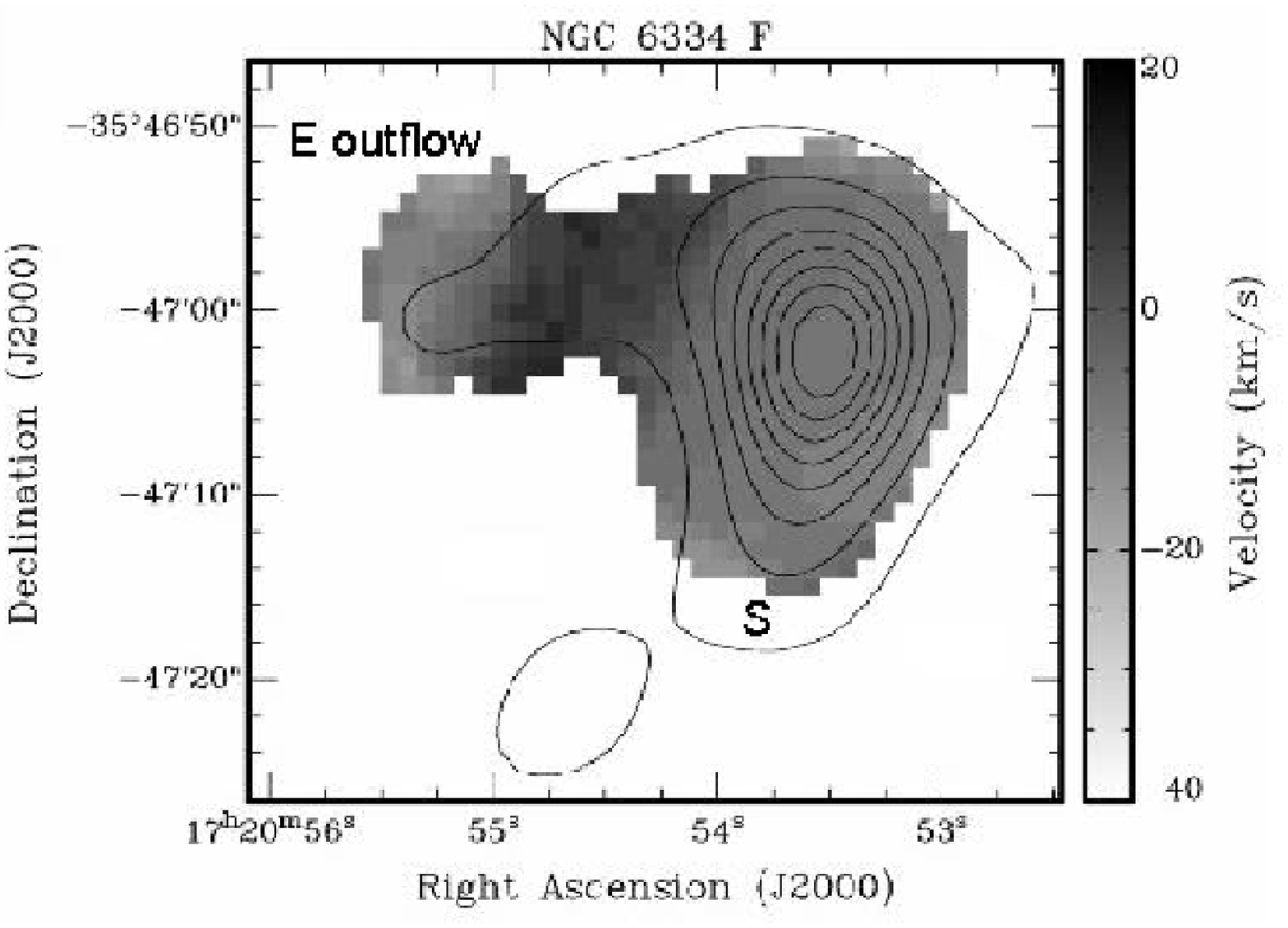}
   \caption{RRL H~91~$\alpha$ moment maps for NGC~6334~F. The plots
            are as in Figure~\ref{fig:Moment318}. Position-velocity
            contours are at 20 to 90\% of maximum emission of
            126.1~mJy~beam$^{-1}$ spaced by 10\%. The lowest contour
            of 25.2~mJy~beam$^{-1}$ corresponds to $4.5
            \sigma$. Continuum contours are at 10 to 90\% of maximum
            emission of 2.13~Jy~beam$^{-1}$ spaced by 10\%.
            Arcsecond-scale emission around the core is seen to the
            east and south of the emission centre. The southern
            component appears at velocities consistent with the -5.33
            km~s$^{-1}$ systemic velocity of the region and represents
            the tail of the cometary region. The eastern component,
            however, peaks at velocities of about +20~km~s$^{-1}$,
            which is significantly different from systemic region
            velocity and the maser velocities of approximately
            -10~km~s$^{-1}$. Therefore, it does not appear to be
            associated with the compact region. This result is
            corroborated by the lack of significant eastern extension
            in the superimposed continuum contours, or the
            high-resolution continuum maps of \otherpaper. This
            emission is in fact known to be coincident with a massive,
            poorly-collimated bipolar outflow
            \protect\cite{BachillerCernicharo90,JacksonEA88,DePreeEA95b}.}
        \label{fig:MomentNGCf}
\end{figure*}

The first of these involved examining cross-cuts through the image
cubes in various position-velocity planes. Positional cross-cuts were
taken to run through the peak brightness position at a range of
angles. Typically, the cuts were made at constant right ascension or
declination or along a line joining the compact region to features of
potential interest.  Part~{\itshape{a}} (left plots) in
Figures~\ref{fig:Moment318}~-~\ref{fig:MomentNGCf} show the
position-velocity plots of integrated flux density for appropriate
cross-cuts in sources exhibiting significant arcsecond-scale extended emission.
Position angle is calculated anticlockwise from a cut in constant declination.
Position is defined as the offset from the plane passing through the
point of maximum H91$\alpha$ emission. If both compact and extended
components are present and physically associated, provided no strong shocks are present,
their systemic velocities should be approximately equal or change smoothly between
the two positions.  Therefore, taking a cut along the line joining
them would result in emission at the same peak velocity, but offset
positionally by the distance between the two components.

Another way of determining whether any two components are likely to be
associated is by plotting the first moment (a flux-weighted velocity mean
across the cube) distribution across the region of H91$\alpha$ emission. 
This is shown in greyscale in part~{\itshape{b}} (right plots) of each
figure. Superimposed are the continuum contours of observations also
made in the 750~D array configuration. 

Clipping levels used in
Fig.~\ref{fig:Moment318}~-~\ref{fig:MomentNGCf} were set to $2 \sigma$, corresponding
to 6~-~11~mJy~beam$^{-1}$ depending on the source (see Table~\ref{tab:ObservedSources}). 
Lower density gas outside the UC \ionhy\/ cores is indicated in the first moment
plots by presence of more emission at velocities close to the peak H91$\alpha$
velocity of the source (given in Table~\ref{tab:GaussianParameters}). In position-velocity plots, this
corresponds to ``peaks'' and/or ``troughs'' (depending on the location
of extended emission) in position at the peak RRL velocity.
Evidence of such emission is seen in G\,318.95$-$0.20
(Figure~\ref{fig:Moment318}), G\,328.81+0.63
(Figure~\ref{fig:Moment328}) and NGC~6334~E
(Figure~\ref{fig:MomentNGCe}) and F~(Figure~\ref{fig:MomentNGCf})
components. Less arcsecond-scale emission is observed around
G\,308.92+0.12, and almost none at all around G\,309.92+0.48. These
results are consistent with the continuum observations of
\otherpaper, and in all cases suggest association between the
compact and more extended components.

Although unlikely, line-of-sight effects cannot be ruled out, as
illustrated by the star-forming complex NGC 6334. The E and F
components of this complex are well-known to be separate star-forming
regions, separated by more than one arcminute on the sky.  However,
their systemic velocities are very similar, with the best fit to the E
component RRL giving a peak velocity of $-4.0 \pm 1.0$ km s$^{-1}$,
compared with $-5.3 \pm 0.6$ km s$^{-1}$ for the F component. For this
reason, the two regions would have been extremely difficult to
distinguish had they been superimposed along our line of sight, rather
than clearly separated on the sky. This suggests that all moment
analysis results should be treated with a degree of caution.

\subsection{Derived Source Parameters}

\subsubsection{Stellar ionising flux and electron temperatures}
\label{sec:ElectronTemps}

Assuming local thermodynamic equilibrium (LTE), the electron
temperature $T_e$ is related to the continuum and
recombination line brightness temperatures $T_C$ and $T_L$, and
the line Full Width at Half Maximum (FWHM) $\Delta V$ in
km~s$^{-1}$, by 
\[ \frac{T_L \Delta V}{T_C} = \frac{6983 \nu^{1.1}}{T_e^{1.15} (1 + Y^+)} \]
where $Y^+$ is the fractional abundance of He$^+$ by number \cite{McGeeNewton91}.
Assuming $Y^+ = 0.1$ and rearranging yields
\begin{equation}
\label{eqn:Te}
        T_e = \left[ \frac{6348\, \nu^{1.1} T_C}{T_L \Delta V} \right]^{0.87}.
\end{equation}

The continuum and recombination line brightness temperatures are
related to the peak flux density $S$ (in Jy) and beam
solid angle $\Omega$ via 
\begin{equation}
\label{eqn:Tbright}
        T_{Bright} = \frac{S \times 10^{-26} c^2}{2 \nu^2 k_B \Omega}.
\end{equation}
The continuum observations were made at $\nu_C =
8.64$~GHz, while the H91$\alpha$ rest frequency is $\nu_L =
8.584$~GHz. $T_e$ can then be determined by combining equations~\ref{eqn:Te} and \ref{eqn:Tbright}. 
In general, optical depth, pressure broadening, and stimulated emission
must be accounted for in RRL analysis \cite{RoelfsemaGoss92}.
These non-LTE effects can be significant for the H91$\alpha$ line.
For expected emission measures of $\sim
10^7$ pc cm$^{-6}$ we use a correction factor of $\frac{T_e}{T_e^*}
\sim 1.3$ \cite{ShaverEA83}. The values of $T_e$ derived above must
be scaled by this factor to account for non-LTE effects.  The resulting
values for $T_e$ are shown in Table~\ref{tab:GaussianParameters}. 
They are somewhat higher than values obtained with single-dish observations
\cite{CaswellHaynes87}. The $T_e$ uncertainties for most sources are
quite large due to uncertainties in the H91$\alpha$ fit parameters.

We also calculate the ionising flux and stellar spectral type for each
source by using the observed continuum fluxes of \otherpaper\/ and
derived electron temperatures via the standard Schraml \&
Mezger~\shortcite{SchramlMezger69} argument. These values are given in
Table~\ref{tab:SourceParameters}, together with the corresponding
spectral types from Panagia~\shortcite{Panagia73}.

\subsubsection{Emission Measures}
\label{sub:EMs}

The continuum brightness temperature $T_C$ is related to electron
temperature $T_e$ via the optical depth $\tau$,
\[ T_C = T_e \left( 1- e^{-\tau} \right) \]
Rearranging this expression, the optical depth is given by
\begin{equation}
\label{eqn:OpticalDepth}
        \tau = -\ln \left( 1 - \frac{T_C}{T_e} \right)
\end{equation}
From equation \ref{eqn:Tbright}, $T_C$ and therefore $\tau$ can be
evaluated for each source using the derived $S_C$ values and beam
solid angles given in Table~\ref{tab:GaussianParameters}.
From this, the peak emission measure for each source can be derived using
\begin{equation}
\label{eqn:EM}
        EM_{peak} = \frac{\tau}{8.235 \times 10^{-2} \alpha(\nu,T_e) T_e^{-1.35} \nu_C^{-2.1}}
\end{equation}
Here, $\alpha(\nu,T_e)$ is a correction factor of order one adopted
from Mezger \& Henderson~\shortcite{MezgerHenderson67}. Values of $\tau$ and peak emission measure
calculated for each source are shown in
Table~\ref{tab:SourceParameters}.

The peak emission measures thus derived were used to estimate the
angular and physical sizes of the ultra-compact components of the
regions. Taking ultra-compact regions to have emission measures in
excess of $10^7$~\pccm, the cutoff emission measure fraction was
determined for each source. This was defined as the lowest contour
level for which emission measure exceeds $10^7$~\pccm, and is
given by {\em cutoff}$ = 10^7/EM_{peak}$.  The location of the closest
contour in the 6~km continuum images of \otherpaper\/ then determined
the size of the observed ultra-compact region, given in
Table~\ref{tab:SourceParameters}. For the purposes of comparison with
our models, the sources G\,308.91+0.12, G\,309.92+0.48 and
G\,318.95$-$0.20 were considered to be spherical; while cometary sources
G\,328.81+0.63 and NGC\,6334F were modeled with the star offset from the
centre of the spherical density distribution.

\begin{table*}
  \caption{Ultra-compact component sizes derived from calculated
    emission measures and continuum contours of \otherpaper, and
    kinematic distances to the sources. Also given are the ionising
    flux for each source, derived from electron temperatures and
    continuum flux densities \protect\cite{SchramlMezger69} of
    Table~\ref{tab:GaussianParameters}, and the corresponding spectral
    type \protect\cite{Panagia73}. The distances to sources are from
    {\itshape{a)}} Phillips \etal\/ \protect\shortcite{PhillipsEA98}; {\itshape{b)}} Caswell
    \& Haynes \protect\shortcite{CaswellHaynes87}; and {\itshape{c)}} Ellingsen
    \etal\/ \protect\shortcite{EllingsenEA96}.  Optical depths and
    peak emission measures are also given where appropriate.  In the
    case of source G\,309.92$+$0.48, Caswell \&
    Haynes~\protect\shortcite{CaswellHaynes87} electron temperature
    scaled for non-LTE effects is used to derive $EM_{peak}$. A scaled
    Caswell \& Haynes~\protect\shortcite{CaswellHaynes87} electron
    temperature is also quoted for source G\,$339.88-1.26$. Continuum
    temperatures for sources G\,328.81$+$0.63 and NGC~6334F are
    comparable with their electron temperatures, and we therefore
    adopt a lower limit of $\tau=1$ for these sources.  Correction
    factors $\alpha$ are interpolated from Mezger \& Henderson
    \protect\shortcite{MezgerHenderson67} values, and range from
    0.9762 to 0.9940.  The cutoff fractions have been determined by
    comparing the derived peak emission measures with the canonical UC
    \ionhy\/ region value of $10^7$~\pccm. Large uncertainties
    associated with calculated $T_e$ and $EM_{peak}$ values arise due
    to propagation of smaller uncertainties through calculations.
    Peak emission measures predicted by the model outlined in
    Section~\ref{sec:model} are also given; these should be treated as
    order-of-magnitude estimates.}
\begin{tabular}{llllrcccccc} 
\hline
 Source & Spectral & log$S_\ast$ & {\scriptsize{Kin. Dist.}} & T$_e$  & $\tau$ &  EM$_{peak}/10^7$ & Cutoff & UC size  & UC size & {\scriptsize{predicted EM$_{peak}/10^7$}} \\
                  & type     & (s$^{-1}$)  & (kpc) & (K)  &  &   (\pccm)     &        & (arcsec) & (pc) & (\pccm)   \\  
\hline
 {\scriptsize{G308.92$+$0.12}} & B\,0 & 47.01 & 5.2$^a$ & {\scriptsize{$8200 \pm 1400$}} & $0.65\pm 0.12$ &  $41 \pm 17$   & 0.024  & 9            & 0.21 & 3.1 \\
 {\scriptsize{G309.92$+$0.48}} & O\,7.5 & 48.52 & 5.3$^a$ & 12200 & $1.54 \pm 0.40$ &         $56 \pm 34$   & 0.018  & 3     & 0.078 & 5.3 \\
 {\scriptsize{G318.95$-$0.20}} & B\,0  & 47.72 & 2.0$^b$ & {\scriptsize{$12600 \pm 800$}} & $1.44 \pm 0.10$ & $56 \pm 8$   & 0.018 & 13    & $0.126$ & 2.6 \\
 {\scriptsize{G328.81$+$0.63}} & O\,8   & 48.39 & 3.0$^b$ & {\scriptsize{$12900 \pm 500$}} & $> 1$ & $41 \pm 4$ & 0.024  & 6 $\times$ 11 & 0.087 $\times$ 0.156 & 4.3 \\
 {\scriptsize{G336.40$-$0.25}} & B\,0.5   & 46.74 & 5.2$^a$ & 4800& - &        -               & -     & -            & -  & - \\
 {\scriptsize{G339.88$-$1.26}} & B\,0.5 & 45.99 & 3.0$^c$ & 10000& - &        -               & -     & -            & - & - \\
 {\scriptsize{G345.01$+$1.79}} & B\,0   & 47.11 & 1.7$^c$ & 10000& - &        -               & -     & -            & - & - \\
 {\scriptsize{NGC 6334F}} & O\,9   & 48.00 & 1.7$^c$ & {\scriptsize{$10100 \pm 500$}}  & $> 1$ &  $30 \pm 7$ & 0.034 & 5 $\times$ 8 & 0.041 $\times$ 0.066 & 4.4 \\
 {\scriptsize{NGC 6334E}} & B\,0$^e$ & 47.13 & 1.7$^c$ & {\scriptsize{$7500 \pm 600$}} & $0.65 \pm 0.07$ & $13 \pm 3$ & 0.079 & - & - & 3.3 \\ 
\hline
  \end{tabular}
  \label{tab:SourceParameters}
\end{table*}

\section{Modeling}
\label{sec:model}

Detailed numerical modeling will certainly be required to address the
nature of extended emission associated with UC \ionhy\/ regions. In
this section, we present a simple, semi-quantitative model. Franco
\etal\/ \shortcite{FrancoEA00a,FrancoEA00b} and Kim \& Koo
\shortcite{KimKoo01} suggest that the ambient density structure is the
primary factor determining \ionhy\/ region sizes and morphologies,
thereby implying the need for more realistic ambient density
representation; e.g. Franco \etal\/ \shortcite{FrancoEA90}. However,
in the present work our focus is on the apparent association between
the ultra-compact components and more diffuse arcsecond-scale extended emission
\cite{WoodChurchwell89a,KurtzEA94}, and we show that this can be explained in
an order of magnitude argument by a hierarchical density model.

The density structure of star-forming cores is an important modeling
parameter. Numerous studies have shown that in low-mass star-forming
clouds the density structure on large scales ($>1$~pc) is well-fit
by power-law distributions $n \propto r^{-p}$.  In high-mass star
formation regions the exponent of the density power-law flattens
significantly for more evolved objects, such as \ionhy\ regions
\cite{BeutherEA02,HatchellvanderTak03,vanderTakEA00}.  High-mass cores
are less well-fit by single power-laws, and show a tendency towards
clumpy substructure, possibly with the clumps embedded within overall
gradients \cite{BeutherEA02,Evans99}.  Given the observational
uncertainty regarding the magnitude of the density gradients and the
scales over which they apply, we have ignored them in our modeling in
favour of a simple approximation of a series of concentric spherical
gas clumps.

We assume that the star forms within a hot core ($R = 0.1$~pc, $T_0
= 200$~K, $n_0 = 10^7$~\cc\/), that is located within a molecular
clump of radius $R = 0.3$~pc, having molecular gas temperature $T_0
= 50$~K and density $n_0 = 10^6$~\cc\/, which itself lies within a
molecular cloud ($R > 0.3$~pc, $T_0 = 25$~K, $n_0 = 10^5$~\cc\/). The
physical characteristics for the interior hot core are taken from
Churchwell \shortcite{Churchwell02}, while those for the intermediate
molecular clump are given by Cesaroni \etal\/ \shortcite{CesaroniEA91}
and Garay \& Lizano \shortcite{GarayLizano99}, and for the exterior
molecular cloud we used the parameters given by Churchwell
\shortcite{Churchwell99}. We note that the values we use to define hot
cores and molecular clumps are indicative only and differ slightly
from those used by Kim \& Koo \shortcite{KimKoo01}.

In the simple model of \ionhy\/ region evolution the radius of the
expanding region is given as a function of time in terms of the initial
Str\"omgren radius $R_s$ and sound speed in the ionised gas $a_i$ by
\cite{DysonWilliams80}
\begin{equation}
        \label{eqn:RdotHierarchy}
        \frac{dR(t)}{dt} = a_i \left(\frac{R(t)}{R_s} \right)^{-3/4}
\end{equation}
The Str\"omgren radius is given as
\[ R_s = \left( \frac{3}{4\pi}\frac{S_\ast}{n_0^2 \beta_2} \right)^{1/3} \]
and assuming the strong-shock limit for the expansion following the 
(instantaneous) formation of the Str\"omgren sphere, we have
\[ R(t) = R_s \left(1+\frac{7 a_i}{4 R_s} t\right)^{4/7} \]

\subsection{Model characteristics}

\subsubsection{Thermal and turbulent pressure}
\label{sec:turb}

Given the higher thermal pressures that we now know to exist in
molecular cores, \ionhy\/ regions produced by O9 or later stars may
still be ultra-compact when they reach pressure equilibrium with their
surroundings \cite{DePreeEA95a}.  The non-thermal broadening of
molecular lines in high-mass star forming regions suggests that
turbulence is present, with velocities of the order of 2~\kms\/
\cite{XieEA96}. The resulting additional turbulent pressure $p_{turb}
= n_0 m_{H_2} V_{turb}^2$, given in terms of the molecular hydrogen
mass $m_{H_2}$ and the turbulent velocity $V_{turb}$ in the
surrounding medium, may act to restrict \ionhy\/ region expansion.

Using equation~\ref{eqn:RdotHierarchy} we can compare the relative
contributions of the expanding ionisation front (I-F) and turbulence in the ambient
medium to the energy balance, $\frac{E_{photo}}{E_{turb}} =
\left( \frac{\frac{dR}{dt}}{V_{turb}} \right)^2 =
\frac{a_i^2}{V_{turb}^2} \left( \frac{R}{R_s} \right)^{-3/2}$

The sound speed in the ionised gas is $a_i=\sqrt{\frac{2 k_B T_e}{m_H}}
\sim 12.9$ km~s$^{-1}$ for an electron temperature of 10\,000~K.
Taking an initial Str\"omgren radius of 0.02~pc, and UC region radius
of 0.1~pc, we have $R = 5 R_s$, and $\frac{E_{photo}}{E_{turb}} \sim 3.7$.
Thus, photoionisation energy  nominally dominates (for $R \approx R_s$) but is
of the same order as the turbulent energy. Turbulent velocities greater than 2~\kms\/
could shift the balance in favour of turbulence. Moreover, as the expansion proceeds,
the I-F energy dominance will die off, as $R$ grows well beyond $R_s$.

\subsubsection{Density structure in ionised regions}
\label{sec:hydrodynamic}

Low-density extended emission on arcminute scales is observed near
many UC \ionhy\/ regions \cite{KurtzEA99,KimKoo01,EllingsenEA05}. By
comparison, as shown in Section~\ref{sub:MomentOutline}, we observe
emission on {\em arcsecond} scales around the UC cores, consistent
with other observations (e.g.  Wood \& Churchwell 1989; Kurtz \etal\/
1994).  Inhomogeneous ambient density structure can explain this
\cite{LiEA05}. Non-uniformity within the ionised region can also arise
if the expansion velocity of the ionisation and shock fronts is
much greater than the sound speed, a condition that occurs early in 
the \ionhy\/ region expansion phase.

The expansion velocity of an \ionhy\/ region slows with time, and is
of the order of the sound speed when the region reaches pressure
equilibrium. The diffusion timescale as the region expands into a
molecular clump is of the order of the sound-crossing time $t \sim
\frac{r_{clump}}{a_i}$. Taking $r_{clump} \sim 0.15$ pc and $a_i \sim
12.9$ km~s$^{-1}$ as before gives $t \sim 1.5 \times 10^4$ years. This
is a significant fraction of an UC \ionhy\/ region lifetime of $\sim
10^5$ years, and hence the ionised gas density cannot be considered
uniform in all cases. This situation is further amplified by the
presence of density inhomogeneities. Clearly, to model \ionhy\/
regions properly, a full hydrodynamical treatment of the problem is
required. Such modeling is beyond the scope of this paper, which
purports only to offer a semi-quantitative plausibility argument.

\subsection{Comparison with observations}
\label{sec:compare}

Apart from NGC\,6334E which happened to be in the same field of view
as NGC\,6334F, the nine regions presented here were selected for the
presence of 6.7-GHz methanol maser emission. These masers are thought
to correspond to a relatively short evolutionary phase that ends soon
after the formation of the \UCHII\/ region (see
\otherpaper). Recombination line analysis of extended emission around
the majority of our sample shows that it is associated with the
compact emission and thus the two must be considered together. We have
compared the predictions from our model (compiled in
Table~\ref{tab:SourceParameters}) with the data for spherical
(G\,308.92+0.12, G\,309.92+0.48 and G\,318.95$-$0.20) and cometary
(G\,328.81+0.63) sources. The cometary source G\,328.81+0.63 was
modeled by positioning the ionising star 0.096~pc from the hot core
centre.  In all cases, observed region sizes agree within a factor of
a few with predicted pressure equilibrium values.  However, the
predicted peak emission measures are consistently more than an order
of magnitude less than the observed values (see
Table~\ref{tab:SourceParameters}). This discrepancy can be explained
by the presence of significant amounts of ionised gas around the UC
region on arcsecond scales, consistent with observational results of
Section~\ref{sub:MomentOutline} and discussed in more detail below.
The remaining sources in our sample, particularly those with complex
morphologies, will require more detailed modeling than is considered
here.

\begin{figure*}
  \centering
  \includegraphics[height=0.4\textwidth]{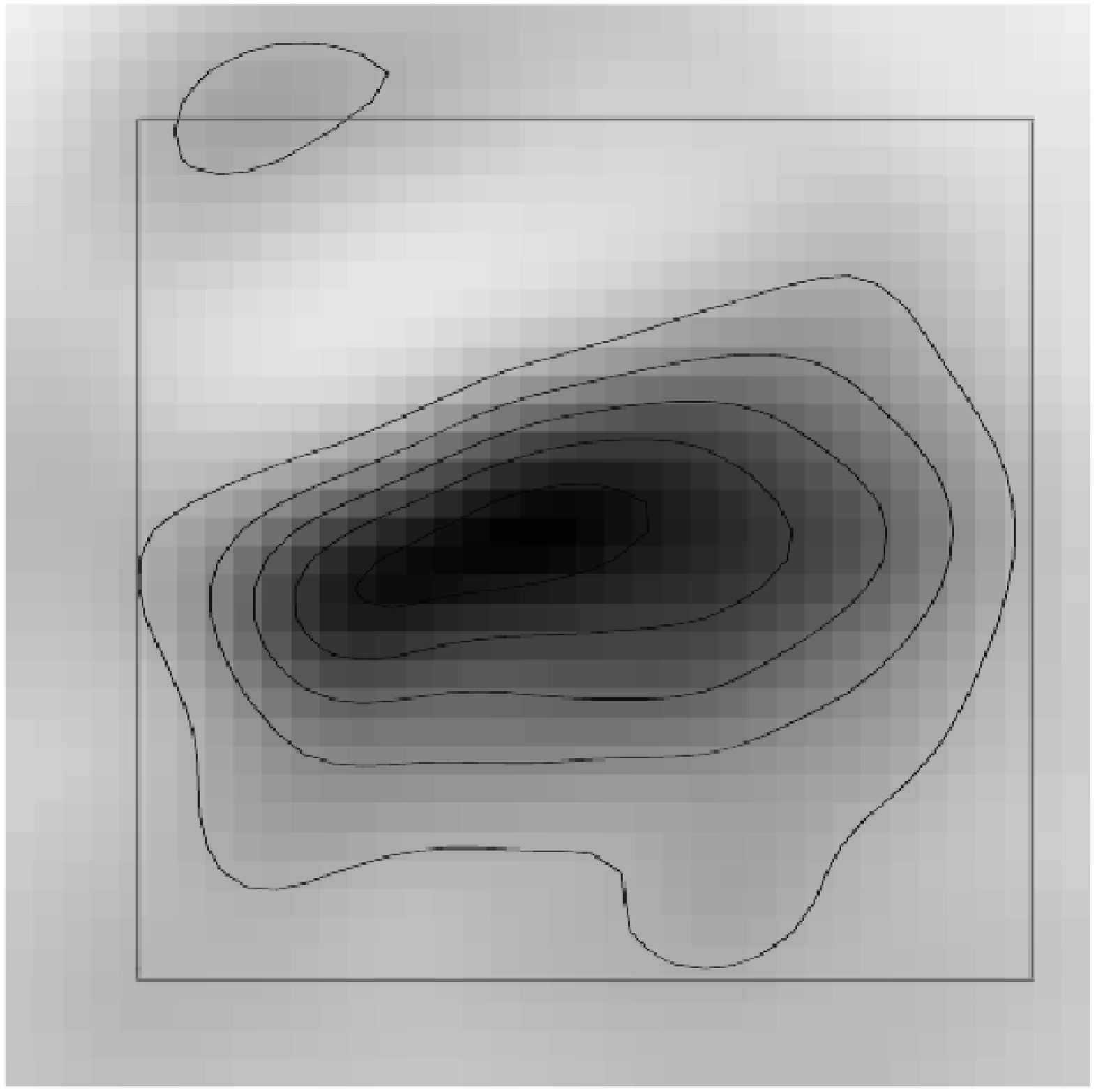}
  \includegraphics[height=0.4\textwidth]{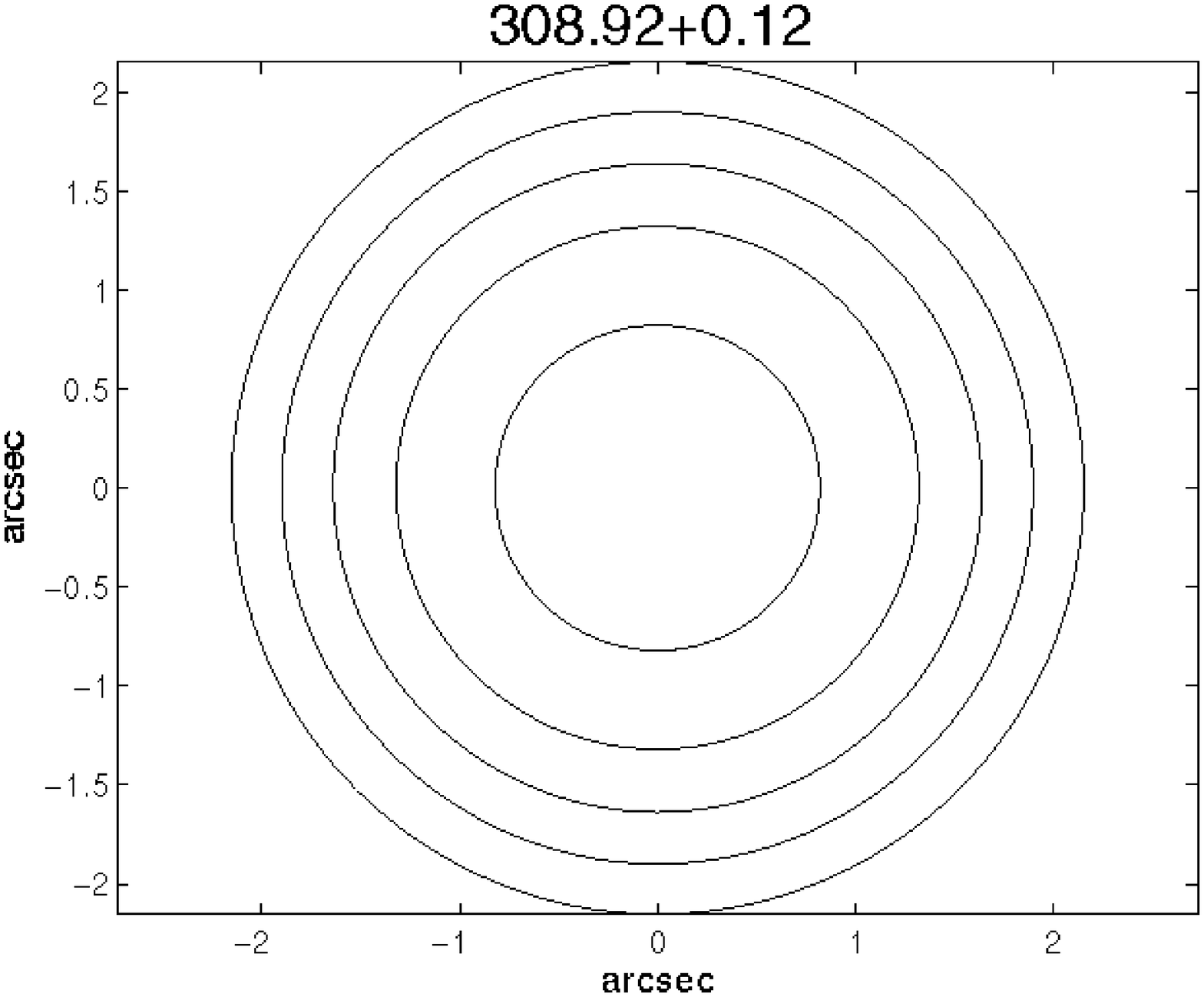}
  \caption{Observed (left) and theoretical (right) emission measure contours for G\,308.92+0.12. 
    Both sets of contours are at 32, 47, 62, 77 and 92\% of the peak
    emission measure. The box plotted together with the observed
    contours encloses the lowest (32\%) theoretical contour. The
    predicted and observed contours are of similar physical size, with
    some evidence for compact core- extended arcsecond-scale emission structure.}
  \label{fig:308contours}
\end{figure*}

For a spherical \ionhy\/ region the distance of a site line from the
centre of the region is $d=r cos \theta$. The
emission measure at this distance is obtained by traversing a length
$2r sin\theta$. For uniform electron density, we then have
$EM(d)/EM_{peak} = \left( 2 n_e^2 r sin\theta \right) / \left( 2 n_e^2
r \right) = sin \theta$. Thus
\begin{equation}
\label{eqn:StromgrenEM}
        \frac{d}{r} = \sqrt{1 - \left( \frac{EM(d)}{EM_{peak}} \right)^2}
\end{equation}
The resultant theoretical contours can then be compared with
observations. The ultra-compact components of sources G\,308.92+0.12,
G\,309.92+0.48 and G\,318.95$-$0.20 were largely unresolved in the
750D array results presented in \otherpaper\/, and high resolution
images made with the 6~km array (also presented in \otherpaper\/) were
used instead. The synthesized beam FWHM was taken as 1.2 arcseconds
for all three sources. For each source, the resulting beam was then
convolved with theoretical contours using the Table
\ref{tab:SourceParameters} ultra-compact region
sizes. Figure~\ref{fig:308contours} shows the theoretical map thus
obtained, together with a high resolution image, for
G\,308.92+0.12. Table~\ref{tab:AxesRatios} shows the theoretical
contour diameter and observed major and minor axes for each contour of
each source. The final three columns give the observed/theoretical
ratios for the two axes, as well as a geometric mean of the two for
sources G\,308.92+0.12 and G\,309.92+0.48. The non-spherical nature of
G\,318.95$-$0.20 (much more so than the other two sources) means that
we have only given the major axis values for this source.

\begin{table*}
  \caption{Theoretical and observed sizes for the three sources
modeled as spherical.  The observed/theoretical major and minor axis
ratios are given for each contour of each source. The geometric means
of the two are also given for sources G\,308.92+0.12 and
G\,309.92+0.48. In source G\,308.92+0.12, the major axis ratio is
almost constant and close to one for most contours, indicating good
agreement between model and observations. By comparison, the major
axis ratios are less and depart significantly from unity in source
G\,318.95$-$0.20, consistent with extended emission observed around
this source (see text). The contours for source G\,309.92+0.48,
although in reasonable agreement with the model, are most likely
unresolved.  The decrease in the observed/theoretical ratio at higher
emission measure contours provides support for existence of ionised
density gradients in the UC cores of all three \ionhy\/ regions. }
  \begin{tabular}{lccccccc} \hline
    Source & EM contour & Theoretical diameter & Major & Minor & Major axis & Minor axis & Geometric \\
         & (\%) & (arcsec) & axis (arcsec) & axis (arcsec) & ratio & ratio & mean ratio \\ \hline
      G\,308.92$+$0.12 & 32 & 4.36 & 4.28 & 2.69 & 0.98 & 0.62 & 0.78 \\
                                                        & 47 & 3.80 & 3.61 & 1.89 & 0.95 & 0.47 & 0.67 \\
                                                        & 62 & 3.29 & 3.11 & 1.42 & 0.95 & 0.43 & 0.64 \\
                                                        & 77 & 2.65 & 2.49 & 0.94 & 0.94 & 0.35 & 0.57 \\
                                                        & 92 & 1.66 & 1.49 & 0.52 & 0.90 & 0.31 & 0.53 \\ \hline 
      G\,309.92$+$0.48 & 11 & - & 2.85 & 2.34 & - & - & - \\
                                                        & 31 & - & 2.05 & 1.72 & - & - & - \\
                                                        & 51 & 2.22 & 1.77 & 1.56 & 0.80 & 0.70 & 0.75 \\
                                                        & 71 & 1.67 & 1.31 & 1.14 & 0.78 & 0.68 & 0.73 \\
                                                        & 91 & 0.93 & 0.56 & 0.47 & 0.60 & 0.51 & 0.55 \\ \hline
      G\,318.95$-$0.20 & 32.5 & 12.58 & 8.22 & - & 0.65 & - & - \\
                                                        & 52.5 & 11.31 & 6.22 & - & 0.55 & - & - \\
                                                        & 72.5 & 9.23 & 3.24 & - & 0.35 & - & - \\
                                                        & 92.5 & 5.08 & 1.25 & - & 0.25 & - & - \\ \hline
  \end{tabular}
  \label{tab:AxesRatios}
\end{table*}

In all three sources the observed/theoretical ratios decrease as we
approach peak emission. Thus the observed contours are slightly denser
near source core, implying the presence of density gradients in the
ionised gas. For G\,309.92+0.48 and especially G\,308.92+0.12 the
ratios are close to constant, suggesting almost uniform region density
and thus that these sources are close to pressure equilibrium. This is
again consistent with a lack of lower density gas observed around
their UC cores. By comparison, the large spread of
observed/theoretical ratios in G\,318.95$-$0.20 suggests a much
steeper ionised gas density gradient in this source, in keeping with
observations of significant arcsecond-scale emission around its UC
core. Accounting for this density gradient would raise the predicted
peak emission measure and thus address the discrepancy between model
and observations discussed in the previous section.

The ratios given in Table~\ref{tab:AxesRatios} are less than one for
all three sources, indicating that the observed \ionhy\/ region sizes
are smaller than model predictions. This could be due to the \ionhy\/
regions not being in pressure equilibrium with the ambient medium --- an
idea consistent with their young ages deduced from maser observations
\cite{EllingsenEA96,PhillipsEA98,DeBuizerEA02}, and also the fact that
this ratio is closer to one for sources G\,308.92+0.12 and
G\,309.92+0.48 which exhibit a more uniform density structure.
G\,309.92+0.48 is unresolved in the 750D array, and this is likely the
main reason for the departure of observed contours from model
predictions. Overestimates of stellar spectral types are another
possible reason for the observed/predicted ratios being less than one,
although this is less likely as radio observations typically
{\em{underestimate}} spectral types due to dust absorption. Other
confinement mechanisms may also play a role. Evidence for non-thermal
broadening in the Gaussian profiles of Figure~\ref{fig:RRLs} lends
further support to this scenario.

The above analysis is applicable to optically thin \ionhy\/ regions.
If we instead had a constant continuum brightness temperature (as
would be expected for an optically thick source), the theoretical
emission measures would be more uniform around the source core,
providing an even greater discrepancy between predictions and
observations.

\section{Discussion}
\label{sec:discussion}

\subsection{Emission Measures}

The fact that some fraction of ionising photons is absorbed by dust
suggests that our measured peak emission measures, which are already
too high to be explained by constant density models, are
underestimates.  This effect can largely be ignored however, as the
attenuation factor is $(1-f)^{1/3}$, where $f$ is the fraction of
photons absorbed by dust \cite{FrancoEA90}, which for $f \sim 0.9$
results in a decrease in the predicted peak emission measure by only a
factor of two.

The large uncertainties associated with the electron temperatures we
derived affect the calculated peak emission measures both directly and
through the value of optical depth $\tau$ in
equation~\ref{eqn:EM}. The logarithmic dependence of $\tau$ on $T_e$
will have a greater effect on the estimated peak emission measure
value than directly through the power-law $T_e$ dependence in the
denominator of equation~\ref{eqn:EM}. Hence the higher electron
temperature estimates we have calculated compared to Caswell \& Haynes
\shortcite{CaswellHaynes87} suggest that, if anything, the estimated
peak emission measures are likely to be {\em{underestimates}}.
Typically, these effects largely cancel each other out, and in any
case affect the derived peak emission measures by at most a factor of
a few.

\subsection{Lifetimes}

The fundamental difference between pressure confinement and other
models is that it predicts that in many cases the observed \ionhy\/
regions are already in equilibrium with their surroundings, rather
than still undergoing expansion. Our modeling suggests pressure
equilibrium may be reached very quickly, with expansion taking place
for only a fraction of the observed lifetimes of \ionhy\/ regions. The
lower \ionhy\/ region age limits thus derived are given in
Table~\ref{tab:Ages}. These are consistent with methanol maser
emission being associated with very young massive stars.

\begin{table*}
  \caption{Lower limits on ultra-compact component ages as predicted by 
the hierarchical ambient density structure model. These are obtained from 
the time required to reach pressure equilibrium for each source.}  
  \begin{tabular}{lc} \hline
    Source & Minimum Age (years) \\ \hline
      G\,308.92$+$0.12 & $2 \times 10^4$ \\
      G\,309.92$+$0.48 & $6 \times 10^4$ \\
      G\,318.95$-$0.20 & $3 \times 10^4$ \\
      G\,328.81$+$0.63 & $5 \times 10^4$ \\
      G\,336.40$-$0.25 & $6 \times 10^3$ \\
      G\,339.88$-$1.26 & $2 \times 10^4$ \\
      G\,345.01$+$1.79 & $5 \times 10^4$ \\
      NGC\,6334F    & $4 \times 10^4$ \\
      NGC\,6334E & $8 \times 10^4$ \\ \hline
  \end{tabular}
  \label{tab:Ages}
\end{table*}

As discussed in section~\ref{sec:turb}, turbulence may provide an
additional confinement mechanism. Any
non-isotropic nature of the turbulence (e.g., if it is
magnetohydrodynamic; Garc\'\i a-Segura \& Franco 1996) may also
contribute to the non-spherical appearance of the resulting \ionhy\/
region. Further investigation of this issue is warranted.

\section{Conclusions}

We have detected arcsecond-scale emission around UC \ionhy\/
cores. Using region parameters derived from continuum and
H\,91$\alpha$ recombination line data we show that although simple
models of expansion in hydrostatic equilibrium reproduce the observed
region sizes, their emission measures are significantly
underestimated. This discrepancy can be explained by the presence of
density gradients in the ionised gas, consistent with young source
ages and observations of the diffuse emission.

\section*{Acknowledgements}

This research has made use of NASA's Astrophysics Data System Abstract
Service. Financial support for this work was provided by the
Australian Research Council.


\begin{thebibliography}{}

\bibitem[\protect\citename{Arthur \& Lizano }1997]{Arthur97}
Arthur, S. J., \& Lizano, S.\ 1997, ApJ, 484, 810

\bibitem[\protect\citename{Bachiller \& Cernicharo }1990]{BachillerCernicharo90}
Bachiller, R. \& Cernicharo, J.\ 1990, A\&A 239, 276

\bibitem[\protect\citename{Beuther \etal\/ }2002]{BeutherEA02}
Beuther, H., Schilke, P., Menten, K.\ M., Motte, F., Sridharan, T.\ K.,
Wyrowski, F. 2002, ApJ 566, 945

\bibitem[\protect\citename{Carral \etal\/ }2002]{CarralEA02}
Carral, P., Kurtz, S. E., Rodr\'\i guez, L. F., Menten, K., Cant\'o, J.,
\& Arceo, R. 2002, AJ, 123, 2574

\bibitem[\protect\citename{Caswell \& Haynes }1987]{CaswellHaynes87}
Caswell, J.\ L., Haynes, R.\ F.\ 1987, A\&A 171, 261

\bibitem[\protect\citename{Cesaroni \etal\/ }1994]{CesaroniEA94}
Cesaroni, R., Churchwell, E., Hofner, P., Walmsley, C.\ M., Kurtz, S.\ 1994,
A\&A 288, 903

\bibitem[\protect\citename{Cesaroni \etal\/ }1991]{CesaroniEA91}
Cesaroni, R., Walmsley, C.\ M., K\"ompe, C., Churchwell, E.\ 1991, 
A\&A 252, 278

\bibitem[\protect\citename{Churchwell }1999]{Churchwell99}
Churchwell, E.\ 1999, in C.\ J. Lada \& N.\ D. Kylafis (eds.),
in {\em {The Origins of Stars and Planetary Systems}}, p. 479, Kluwer, 
Dordrecht

\bibitem[\protect\citename{Churchwell }2002]{Churchwell02}
Churchwell, E.\ 2002, ARA\&A 40, 27

\bibitem[\protect\citename{De~Buizer \etal\/ }2002]{DeBuizerEA02}
De~Buizer, J.\ M., Walsh, A.\ J., Pi\~{n}a, R.\ K., Phillips, C.\ J., Telesco, C.\ M. 2002, ApJ 564, 327

\bibitem[\protect\citename{De~Pree \etal\/ }1995a]{DePreeEA95a}
De~Pree, C.\ G., Rodr\'\i guez, L.\ F., Goss, W.\ M.\ 1995a, Rev.~Mex.~A\&A 31, 39

\bibitem[\protect\citename{De~Pree \etal\/ }1995b]{DePreeEA95b}
De~Pree, C.\ G., Rodr\'\i guez, L.\ F., Dickel, H.\ R., Goss, W.\ M. 1995b, ApJ 447, 220

\bibitem[\protect\citename{Dyson \& Williams }1980]{DysonWilliams80}
Dyson, J.\ E., Williams, D.\ A.\ 1980,
{\em {Physics of the Interstellar Medium}}, Wiley, London

\bibitem[\protect\citename{Dyson, Williams \& Redman }1995]{DysonEA95}
Dyson, J.\ E., Williams, R.\ J.\ R., Redman, M.\ P. 1995, MNRAS 277, 700

\bibitem[\protect\citename{Ellingsen \etal\/ }1996]{EllingsenEA96}
Ellingsen, S.\ P., Norris, R.\ P., McCulloch, P.\ M.\ 1996, MNRAS 279, 101

\bibitem[\protect\citename{Ellingsen \etal\/ }2003]{EllingsenEA03}
Ellingsen, S.\ P., Cragg, D.\ M., Minier, V., Muller, E., Godfrey, P.\ D., 
2003, MNRAS 344, 73 

\bibitem[\protect\citename{Ellingsen, Shabala \& Kurtz }2005]{EllingsenEA05} 
Ellingsen, S.\ P., Shabala, S.\ S., Kurtz, S.\ E.\ 2005, MNRAS 357, 1003 (\otherpaper)

\bibitem[\protect\citename{Evans \etal\/ }1999]{Evans99}
Evans, N.\ J. II 1999, Annual Review of Astronomy and Astrophysics 37, 311

\bibitem[\protect\citename{Feldt \etal\/ }2004]{FeldtEA04} 
Feldt, M., Puga, E., Lenzen, R., Henning, Th., Brandner, W., Stecklum, B.,
Langrage, A.\ M., Gendron, E., Rousset, G.\ 2004, ApJ in press

\bibitem[\protect\citename{Felli \etal\/ }1997]{FelliEA97}
Felli, M., Testi, L., Valdettaro, R., Wang, J.-J. 1997, ApJ 484, 375

\bibitem[\protect\citename{Franco \etal\/ }1990]{FrancoEA90}
Franco, J., Tenorio-Tagle, G., Bodenheimer, P.\ 1990, ApJ 349, 126

\bibitem[\protect\citename{Franco \etal\/ }2000a]{FrancoEA00a}
Franco, J., Kurtz, S.\ E., Garc\'\i a-Segura, G., Hofner, P.\ 2000a, Ap\&SS 272, 169

\bibitem[\protect\citename{Franco \etal\/ }2000b]{FrancoEA00b}
Franco, J., Kurtz, S., Hofner, P., Testi, L., Garc\'\i a-Segura, G., 
Martos, M.\ 2000b, ApJ 542, L143

\bibitem[\protect\citename{Garay \& Lizano }1999]{GarayLizano99}
Garay, G. \& Lizano, S.\ 1999, PASP 111, 1049

\bibitem[\protect\citename{Garc\'{\i}a-Segura \& Franco }1996]{GarciaSeguraFranco96}
Garc\'\i a-Segura, G., Franco, J.\ 1996, ApJ 469, 171

\bibitem[\protect\citename{Hatchell \& van der Tak }2003]{HatchellvanderTak03}
Hatchell, J., van der Tak, F.\ F.\ S. 2003, A\&A 409, 589 

\bibitem[\protect\citename{Hollenbach \etal\/ }1994]{HollenbachEA94}
Hollenbach, D., Johnstone, D., Lizano, S., Shu, F. 1994, ApJ 428, 654

\bibitem[\protect\citename{Icke }1979]{Icke79}
Icke, V. 1979, A\&A 78, 352

\bibitem[\protect\citename{Jackson \etal\/ }1988]{JacksonEA88}
Jackson, J.\ M., Ho, P.\ T.\ P., Haschick, A.\ D. 1988, ApJ 333, L73

\bibitem[\protect\citename{Keto }2003]{Keto03}
Keto, E. 2003, ApJ 599, 1196

\bibitem[\protect\citename{Kim \& Koo }1996]{KimKoo96}
Kim, K.-T. \& Koo, B.-C.\ 1996, Journal of the Korean Astronomical Society Supp. 29, S177

\bibitem[\protect\citename{Kim \& Koo }2001]{KimKoo01}
Kim, K.-T., Koo, B.-C.\ 2001, ApJ 549, 979

\bibitem[\protect\citename{Kim \& Koo }2002]{KimKoo02}
Kim, K.-T., Koo, B.-C.\ 2002, ApJ 575, 327

\bibitem[\protect\citename{Koo \& Kim }2003]{KooKim03}
Koo, B.-C., Kim, K.-T.\ 2003, ApJ 596, 362 

\bibitem[\protect\citename{Kurtz \etal\/ }1994]{KurtzEA94}
Kurtz, S.\ E., Churchwell, E., Wood, D.\ O.\ S.\ 1994, ApJS 91, 659

\bibitem[\protect\citename{Kurtz \etal\/ }1999]{KurtzEA99}
Kurtz, S.\ E., Watson, A.\ M., Hofner, P., Otte, B.\ 1999, ApJ 514, 232

\bibitem[\protect\citename{Li \etal\/ }2005]{LiEA05}
Li, Y., MacLow M.-M., Abel, T.\ 1999, ApJ 610, 339  
 this is a radiative transfer paper, not hydro simulations!

\bibitem[\protect\citename{MacLow \etal\/ }1991]{MacLowEA91}
MacLow, M.-M., van Buren, D., Wood, D.\ O.\ S., Churchwell, E.\ 1991,
ApJ 369, 395

\bibitem[\protect\citename{McGee \& Newton }1981]{McGeeNewton91}
McGee, R.\ X., Newton, L.\ M.\ 1981, MNRAS 196, 889

\bibitem[\protect\citename{Mezger \& Henderson }1967]{MezgerHenderson67}
Mezger, P.\ G., Henderson, A.\ P.\ 1967, ApJ 147, 471

\bibitem[\protect\citename{Panagia }1973]{Panagia73}
Panagia, N.\ 1973, AJ 78, 929

\bibitem[\protect\citename{Panagia \etal\/ }1978]{PanagiaEA78}
Panagia, N., Natta, A., Preite-Martinez, A.\ 1978, A\&A 68, 265

\bibitem[\protect\citename{Phillips \etal\/ }1998]{PhillipsEA98}
Phillips, C.\ J., Norris, R.\ P., Ellingsen, S.\ P., McCulloch, P.\ M.\ 1998,
MNRAS 300, 1131

\bibitem[\protect\citename{Schraml \& Mezger }1969]{SchramlMezger69}
Schraml, J., Mezger, P.\ G.\ 1969, ApJ 156, 269


\bibitem[\protect\citename{Spitzer }1978]{Spitzer78}
Spitzer, L.\ 1978, {\em{Physical Processes in the Interstellar Medium}}, Wiley, New York

\bibitem[\protect\citename{Rodr\'\i guez \etal\/ }1982]{RodriguezEA82}
Rodr\'\i guez, L.\ F., Canto, J., Moran, J.\ M.\ 1982, ApJ 103, 110

\bibitem[\protect\citename{Roelfsema \& Goss }1992]{RoelfsemaGoss92}
Roelfsema, P.\ R., Goss, W.\ M. 1992, A\&A Rev. 4, 161

\bibitem[\protect\citename{Shaver \etal\/ }1983]{ShaverEA83}
Shaver, P.\ A., McGee, R.\ X., Newton, L.\ M., Danks, A.\ C., Pottasch, 
S.\ R.\ 1983, MNRAS 204, 53

\bibitem[\protect\citename{Tenorio-Tagle }1979]{Tenorio-Tagle79}
Tenorio-Tagle, G. 1979, A\&A 71, 59

\bibitem[\protect\citename{Testi \etal\/ }1995]{TestiEA95}
Testi, L., Olmi, L., Hunt, L., Tofani, G., Felli, M., Goldsmith, P. 1995, A\&A 303, 881

\bibitem[\protect\citename{van Buren \etal\/ }1990]{vanBurenEA90}
van Buren, D., MacLow, M.-M., Wood, D.\ O.\ S., Churchwell, E.\ 1990,
ApJ 353, 570

\bibitem[\protect\citename{van der Tak \etal\/ }2000]{vanderTakEA00}
van der Tak, F.\ F.\ S., van Dishoeck, E.\ F., Evans, N.\ J. II, Blake, G.\ A.
2000, ApJ 537, 283

\bibitem[\protect\citename{Walsh \etal\/ }1998]{WalshEA98}
Walsh, A.\ J., Burton, M.\ G., Hyland, A.\ R., Robinson, G.\ 1998,
MNRAS 301, 640

\bibitem[\protect\citename{Wood \& Churchwell }1989]{WoodChurchwell89a}
Wood, D.\ O.\ S., Churchwell, E.\ 1989, ApJS 69, 831

\bibitem[\protect\citename{Wood \etal\/ }2005]{WoodEA05}
Wood, K., Haffner, L.\ M., Reynolds, R.\ J., Mathis J.\ S., Madsen, G.\ 2005, ApJ 633, 295

\bibitem[\protect\citename{Xie \etal\/ }1996]{XieEA96}
Xie, T., Mundy, L.\ G., Vogel, S.\ N., Hofner, P.\ 1996, ApJ 473, L131

\end{thebibliography}
\end{document}